\documentclass[%
 amsmath,amssymb,
 preprint,
 aps,
prl,
]{revtex4-1}
\usepackage{graphicx,pstricks}
\usepackage{graphics}
\usepackage{dcolumn}
\usepackage{bm}
\usepackage{amsmath}

\usepackage[percent]{overpic}
\usepackage{array}
\newcolumntype{P}[1]{>{\centering\arraybackslash}p{#1}}

\newcommand{\beginsupplement}{%
        \setcounter{table}{0}
        \renewcommand{\thetable}{S\arabic{table}}%
        \setcounter{figure}{0}
        \renewcommand{\thefigure}{S\arabic{figure}}%
     }
 
\usepackage[colorinlistoftodos]{todonotes}

\begin{document}


\title{Mechanical control of crystal symmetry and superconductivity in Weyl semimetal MoTe$_2$ }


\author{
Colin Heikes,$^1$
I-Lin Liu,$^{1,2,4}$
Tristin Metz,$^2$
Chris Eckberg,$^2$
Paul Neves,$^2$,
Yan Wu,$^3$ 
Linda Hung,$^1$
Phil Piccoli, $^5$
Huibo Cao,$^3$ 
Juscelino Leao,$^{1}$,  
Johnpierre Paglione,$^2$
Taner Yildirim,$^1$ 
Nicholas P. Butch,$^{1,2}$ and William Ratcliff II$^1$}

\affiliation{\vspace{4pt}$^1$NIST center for Neutron Research, NIST, Gaithersburg, Maryland 20899, USA.
\\$^2$Center for Nanophysics and Advanced Materials, University of Maryland, College Park, Maryland 20742, USA.
\\$^3$Neutron Scattering Division, Oak Ridge National Laboratory (ORNL), Oak Ridge, Tennessee 37831, USA.
\\$^4$Department of  Materials Science and Engineering, University of Maryland, College Park, MD 20742 USA.
\\$^5$Department of Geology, University of Maryland, College Park, MD 20742 USA.}
 \email[Corresponding Author:]{william.ratcliff@nist.gov}

\date{\today}

\begin{abstract}
The non-centrosymmetric Weyl semimetal candidate,  MoTe$_2$ was investigated through neutron diffraction and transport measurements at pressures up to 1.5 GPa and at temperatures down to 40 mK.  Centrosymmetric and non-centrosymmetric structural phases were found to coexist in the superconducting state.   Density Functional Theory (DFT) calculations reveal that the strength of the electron-phonon coupling is similar for both crystal structures.  Furthermore, it was found that by controlling non-hydrostatic components of stress, it is possible to mechanically control the ground state crystal structure.  This allows for the tuning of crystal symmetry in the superconducting phase from centrosymmetric to non-centrosymmetric.   DFT calculations support this strain control of crystal structure. This mechanical control of crystal symmetry gives a route to tuning the band topology of MoTe$_2$ and possibly the topology of the superconducting state.  

 \end{abstract}

\maketitle



Topological superconductivity, which arises when a bulk superconducting state coexists with a topologically non-trivial band structure, leading to gapless surface states in a superconducting system, is of particular interest and excitement due to the possibility of stabilizing exotic Majorana excitations\cite{sato2017topological}. One promising route to realizing topological superconductivity is finding superconductivity in materials with topologically non-trivial band structures, as is found in semimetallic MoTe$_2$, where both a type II Weyl Semimetallic state and a superconducting state have been reported \cite{qi2016superconductivity,luo2016toposuper,guguchia2017uSR,huang2016spectroscopic,deng2016experimental,deng2017revealing,rhodes2016impurity}. This type II Weyl semimetallic state is enabled by an inversion symmetry breaking structural transition which takes place at a transition temperature (T$_S$) around 250 K \cite{clarke1978,brown1966crystal,takahashi2017anticorrelation}. The superconductivity, the topology, and structure of MoTe$_2$ have been demonstrated to be strongly influenced by both doping \cite{rhodes2016impurity,cho2017te,rhodes2017Wengineering,chen2016Sdoping,belopolski2016discovery,lv2017composition,oliver2017structural,takahashi2017anticorrelation} and pressure \cite{qi2016superconductivity,guguchia2017uSR,takahashi2017anticorrelation}. Interestingly, pressure and doping increase the superconducting transition temperature (T$_c$) while apparently reducing T$_S$, though the coupling between the electronic ground state and the crystal structure is an open question. Here we study the effect of pressure on both superconductivity and the observed structural phase transition in detail and  show that the deliberate application of pressure in hydrostatic or non-hydrostatic form allows us to control the crystal symmetry in this material and thus gives us a route to tuning the topology of the superconducting state. 

The proposed type-II Weyl semimetal and superconductor MoTe$_2$ offers the opportunity for realizing topological superconductivity through the coexistence of a topologically nontrivial band structure with superconductivity. An open question in this material is the nature of the interplay between pressure, the electronic ground state, and the structural transition between a centrosymmetric monoclinic structure (the 1T' phase) and a non-centrosymmetric orthorhombic structure (the T$_d$ phase). We show through a combination of temperature and pressure dependent transport and elastic neutron scattering measurements that the two possible structures can coexist at a range of pressures and temperatures concurrent with superconductivity. We also illustrate that the ground state crystal structure can be controlled independent of the superconductivity through non-hydrostatic stress, allowing for a centrosymmetric superconducting state, a non-centrosymmetric superconducting ground state, or a superconducting mixed structure state. Our Density Functional Theory (DFT) calculations illustrate the near degeneracy of the two structural phases as well as the small energy barrier between phases, explaining our observation of a mixed phase state under hydrostatic pressure conditions. Unlike the typical case of inversion symmetry breaking structural transitions in perovskite ferroelectrics or geometrically designed polar metals \cite{kim2016polar,benedek2013there}, we also show that this structural transition is driven not by a phonon mode softening to an imaginary vibrational frequency as is suggested in \cite{duerloo2014structural}, but rather due to entropic considerations. Our calculations illustrate that the pressure dependent superconductivity in MoTe$_2$ can be reproduced from single layer simulations, consistent with the decoupling of ground state structure and superconductivity. Further, our calculations offer justification for why non-hydrostatic stresses alter the ground state crystal structure and allow for selection between centrosymmetric and non-centrosymmetric states.\par
\begin{figure}
\begin{center}
\includegraphics[width=\columnwidth]{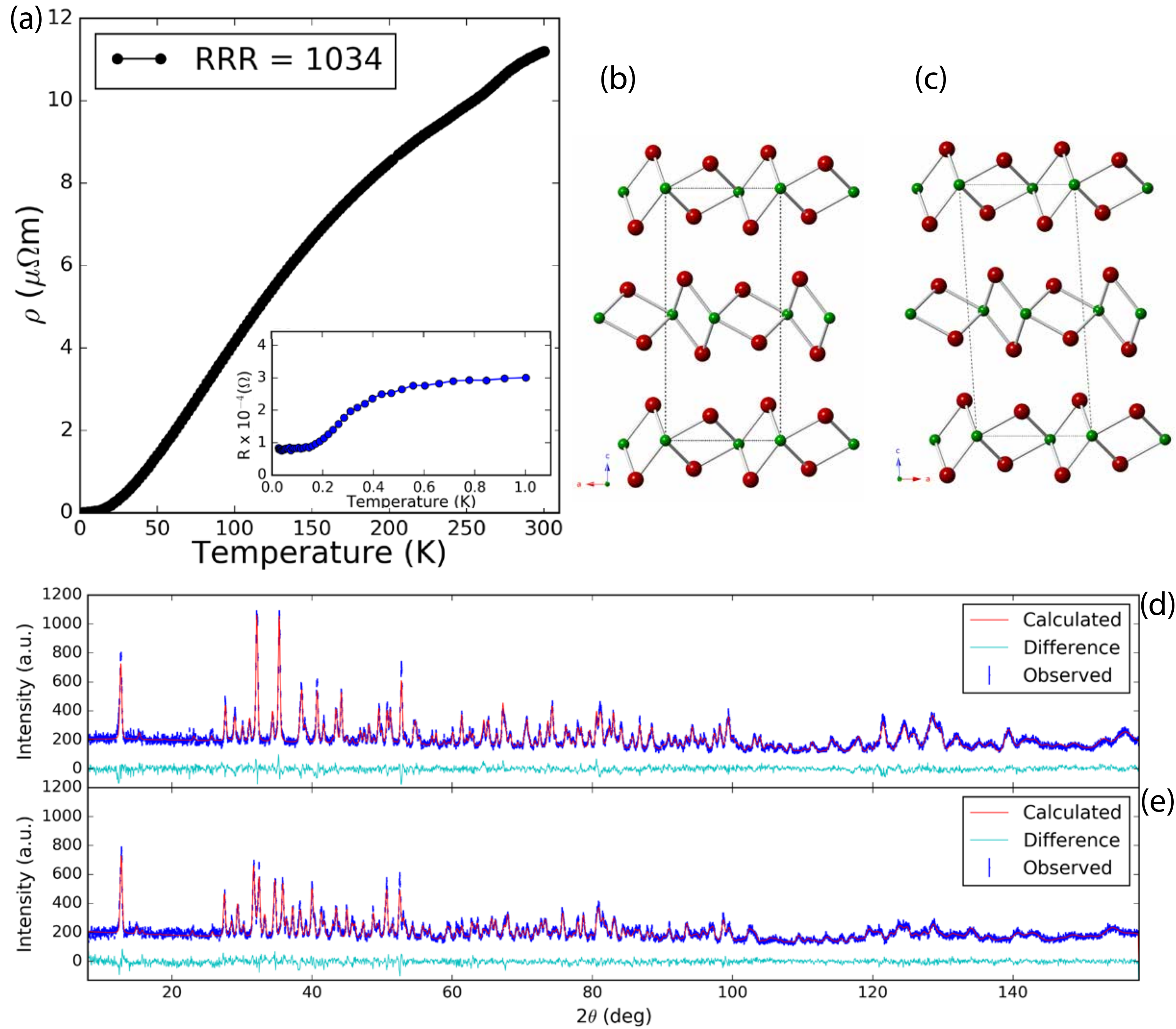}
\end{center}
\caption[Structure and Transport of MoTe$_2$ Single Crystals]{Structure and Transport of MoTe$_2$ Single Crystals.  (a). Temperature dependent longitudinal resistivity of a single crystal with a RRR value of 1034 typical of our synthesis. The inset illustrates the second turnover and non-zero saturation of the resistivity below 1 K indicative of the onset of incomplete superconductivity. (b).,(c). Crystal structure of the T$_d$ (b) and 1T' (c) phases of MoTe$_2$ illustrating the shear displacement of the unit cell. (d).,(e). Reitveld refined neutron powder diffraction measurements of MoTe$_2$ at 3 K  in the T$_d$ phase (e). and at 300 K in the 1T' phase (f). Powder fit parameters and refinement statistics are shown in tables S1-3. }  \label{fig1}
\end{figure}

We have performed temperature dependent longitudinal resistivity measurements as well as longitudinal magnetoresistance measurements on a variety of crystals from multiple batches as described in the supplement \cite{supplement1}. The results for a typical crystal are shown in figure \ref{fig1}(a). We clearly see the transport anomaly associated with the structural transition (at $T_S$) from the monoclinic 1T' phase to the T$_d$ phase \cite{qi2016superconductivity,rhodes2016impurity,zandt2007quadratic,hughes1978electrical}. This particular crystal shows a RRR value (defined as the ratio of the resistance at 300 K to the resistance at 2 K) of \textgreater 1000 as well as a MR ratio of \textgreater 190,000\% at 2 K and 15 T which illustrates the high sample quality. From electron probe microanalysis/wavelength-dispersive x-ray spectroscopy(EPMA-WDX), we measure that our crystals have stoichiometric composition within our measurement error, with no obvious trends in RRR value with sample composition and no apparent composition gradients within a given crystal. As is shown in the inset of figure \ref{fig1} (a), this sample also has a resistive turnover above a temperature of 0.4 K indicating the onset of superconductivity (T$_c$), which is consistent with the sample quality dependent superconductivity reported in \cite{rhodes2016impurity}. However, we do not see a full transition to a zero resistance state at ambient pressure down to 25 mK in contrast with previous reports of superconductivity in this system \cite{guguchia2017uSR,qi2016superconductivity,rhodes2016impurity,takahashi2017anticorrelation}. Partial volume fraction superconductivity is confirmed with single crystal ac-susceptibility measurements illustrating the onset of superconductivity with small volume fractions at ambient pressure.  \par

\begin{figure}
\begin{center}
\includegraphics[width=\columnwidth]{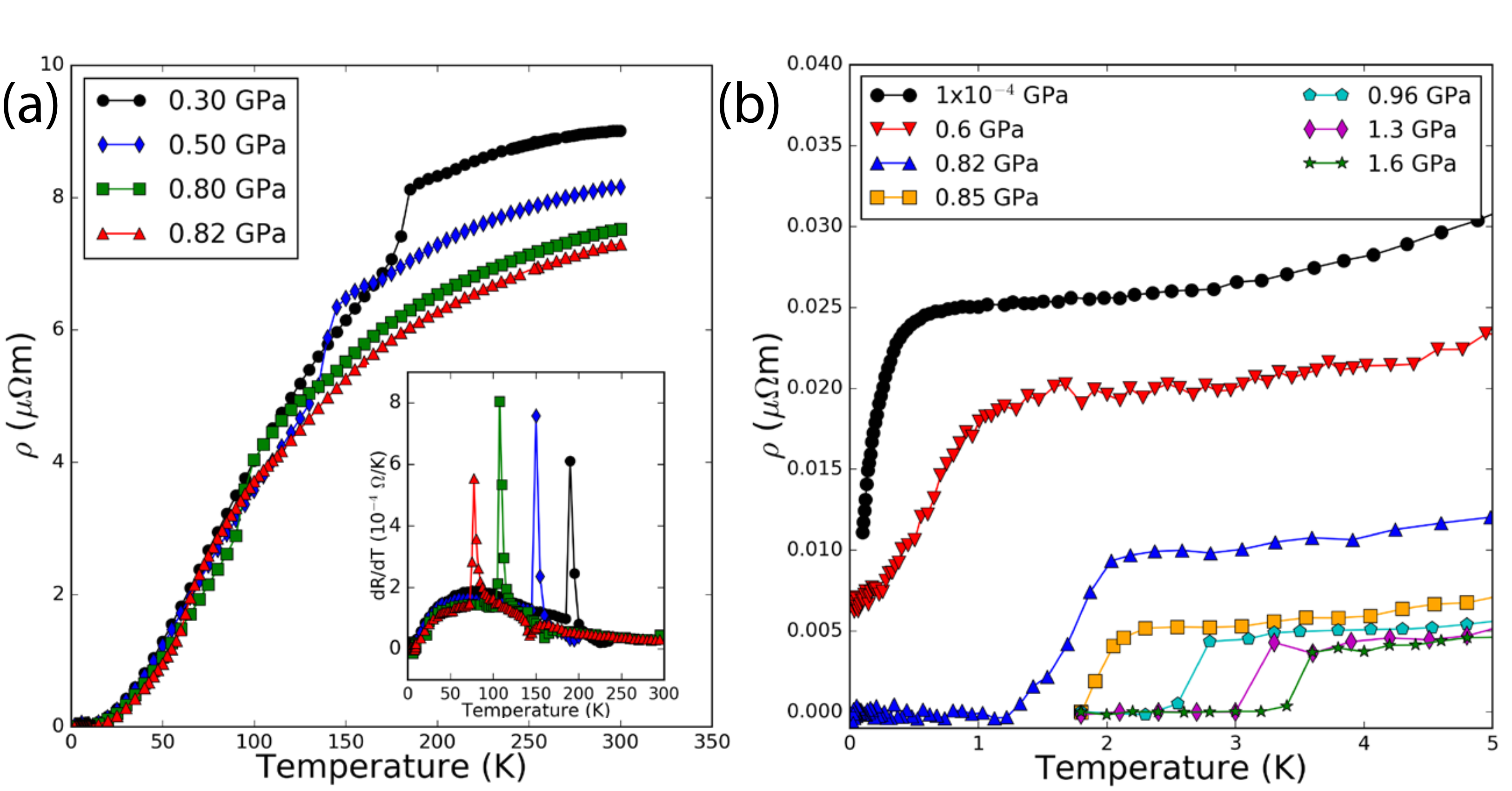}
\end{center}
\caption[Pressure Dependence of Transport Measurements]{Pressure Dependence of Transport Measurements. (a). Pressure dependent resistivity upon heating from 1.5 K. The kink in the resistivity indicates the position of the structural transition from the T$_D$ phase to the 1T' phase. Inset shows differential resistance vs. temperature clearly indicating T$_S$. We no longer see evidence of T$_s$ above 0.82 GPa. (b). Pressure dependence of the superconducting T$_c$. We see a full resistive transition at 0.82 GPa and above.   }  \label{fig2}
\end{figure}

The relationship between pressure enhancement of superconductivity and the pressure driven transition to the 1T' state from the T$_d$ state in both MoTe$_2$ and WTe$_2$ have been taken as evidence of a relationship between the structural transition and T$_c$ enhancement \cite{takahashi2017anticorrelation,lu2016origin} though this is not a settled matter in either material \cite{zhou2016pressure,guguchia2017uSR,takahashi2017anticorrelation}. Via transport measurements, we are able to track a suppression of the T$_d$ phase with pressure up to 0.82 GPa where T$_S$ is suppressed to below 80 K as is shown in figure 2(a). Further increases in pressure show no obvious kink in the resistivity nominally indicating that the non-centrosymmetric phase is unstable above 0.8 GPa, in contrast with the pressure phase diagram in \cite{qi2016superconductivity} but consistent with the reports of \cite{takahashi2017anticorrelation} where the crystal symmetry change is assumed to enhance superconductivity. Furthermore, this 0.82 GPa pressure is also the point at which we observe the transition from a partially superconducting state to a full zero resistance state as is shown in figure \ref{fig2}(b). \par

Since the Weyl semimetal state can only exist with broken inversion symmetry, it is critical to directly probe the crystal structure of MoTe$_2$ in the superconducting state.  Using elastic neutron scattering we have probed the 1T' to T$_d$ structural transition as a function of pressure and temperature. To do this, we selected one set of reflections distinct between the T$_d$ and 1T' phases in one crystal zone (the (201) like reflections) and one set of reflections common to both phases  (the (008) reflections) in the same zone and monitored those reflections through phase space. The convention for labeling (hkl) and crystallographic a, b, and c axes in the T$_d$ and 1T' phases varies in the literature. Our convention for axis labeling and our reflection choice is explained in supplement section II. A \cite{supplement3}.  We will refer to the distinct reflections as the monoclinic (coming from the 1T' phase fraction) and orthorhombic (coming from the T$_d$ phase fraction) reflections while referring to the common reflections as the (00l) reflections. Details of the various neutron scattering measurements can be found in the experimental methods section \cite{supplement1}.\par
 At ambient pressure, we clearly see a first order transition from the 1T' to the T$_d$ phase upon cooling from room temperature while monitoring both the monoclinic and orthorhombic reflections, with a large coexistence region of more than 50 K. The mixed phase state is stable at these temperatures for timescales on the order of hours. Upon heating, we observe the return to the 1T' phase, though we observe a much larger coexistence region than is seen from transport. Our coexistence region is in line with previous Raman measurements and x-ray measurements which show a coexistence region of \textgreater50 K and the survival of a mode attributed to the T$_d$ phase up to room temperature upon warming from the T$_d$ phase \cite{chen2016RAMAN,clarke1978}. This suggests that the transport signature, while clearly linked with the structural transition, is not a direct measure of the structural transition volume fraction. Instead, it may be indicating a percolation-like transition upon cooling (warming) with increasing (decreasing) T$_d$ phase fraction. We also note that we see equal monoclinic twin populations both in the as grown samples and after cycling through the phase transition. \par
 
We next cooled our crystal down to 40 mK and confirmed that we saw no evidence of any reentrant monoclinic phase transition upon the onset of superconductivity. We also performed reciprocal space maps at a range of temperatures between 40 mK and 2 K, and see no evidence of any modulation of the intensity or shape of the orthorhombic reflections as the sample crosses the measured T$_c$ for partial superconductivity. Despite our observation that our crystals do not reach a zero resistance state by 25 mK, if superconductivity were confined to monoclinic sample regions we would have expected to see a monoclinic phase fraction in the scattering.  \par

\begin{figure}
\begin{center}
\includegraphics[width=\columnwidth]{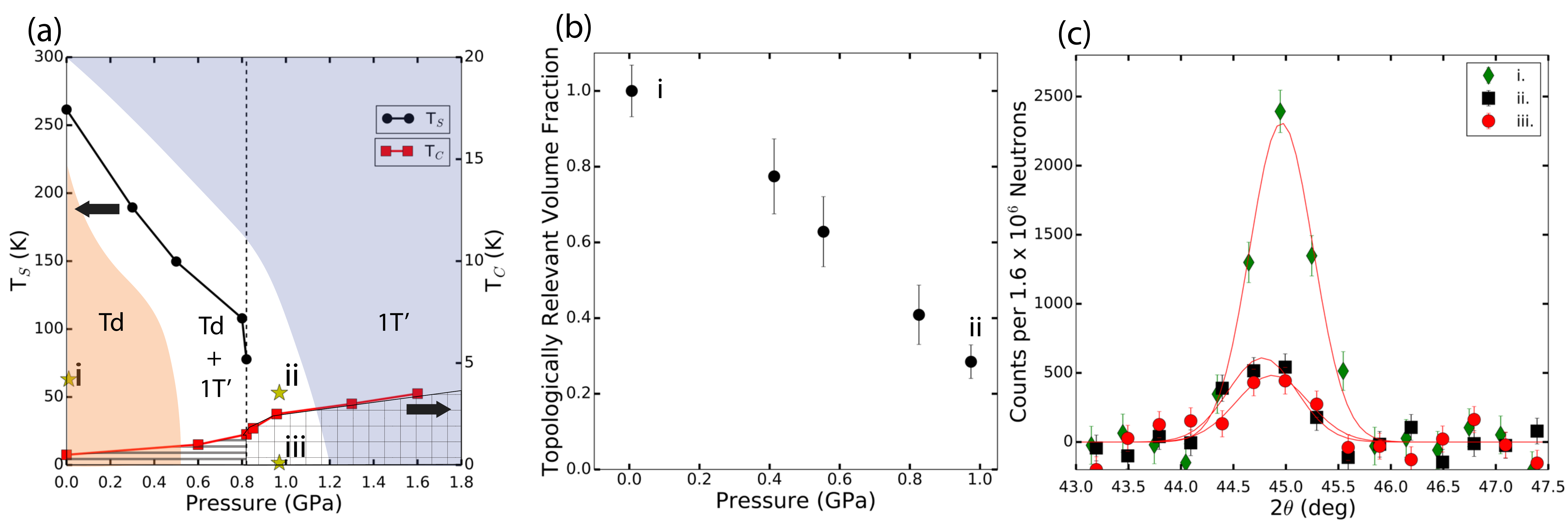}
\end{center}
\caption[Phase Diagram and Pressure Dependent Neutron Scattering for MoTe$_2$]{Phase Diagram and Pressure Dependent Neutron Scattering for MoTe$_2$. (a). Transport generated phase diagram. Black circles represent 1T' to T$_d$ structural transition temperature obtained from the dR/dT upon warming, red squares indicate onset of superconductivity from dR/dT. The dotted vertical line indicates the pressure at which we see concurrent loss of a structural resistance signature as well as the onset of a full zero resistance state. The yellow stars labeled with lower case roman numerals indicate the neutron measurements shown in (b) and (c). Horizontal cross hatching indicates partial superconductivity and grid cross hatching indicates full resistive transitions. Background color indicates structural phase (b). Phase fraction of the T$_d$ phase as a function of applied pressure measured at 63 K. (c). Longitudinal scans along the orthorhombic peaks at points i, ii, and iii on the phase diagram in (a). Data is background subtracted. } \label{fig3}
\end{figure}

Using a steel based gas pressure cell compatible with \textit{in situ} neutron scattering as described in the supplement \cite{supplement1} and illustrated in figure 4(a), we monitored the same orthorhombic and monoclinic reflections as well as an (008) reflection over a pressure range from 0.02 GPa to 1 GPa in a temperature range from 1.5 K to 100 K. We initial cooled our sample to 63 K at 0.02 GPa and confirmed the expected T$_d$ structure at this phase point (point i in figure \ref{fig3} A). The 63 K temperature is chosen to maintain the He pressure medium in a liquid or gaseous state over the entire pressure range up to 1 GPa.  We then increased the pressure by supplying more He gas, and monitored the integrated intensity of longitudinal scans at the orthorhombic position. For these neutron measurements, all error bars and confidence intervals are given by standard deviations of the Poisson distribution. \par 
Upon pressure increase, we immediately observe the start of the transition from the T$_d$ to the 1T' phase, but surprisingly we see that a significant phase fraction (30$\pm$5 \%) of the T$_d$ phase survives up to our maximum pressure of 1 GPa, which is well above the nominal critical pressure from transport \cite{takahashi2017anticorrelation}. This pressure dependence of the T$_d$ phase fraction is shown in figure \ref{fig3}(b), where the extracted phase fractions come from the ratio of the integrated intensity of the orthorhombic reflection (labeled as the (201)$_O$  reflection) at a given temperature and pressure to the intensity at 0.02 GPa and 63 K where we have full volume fraction T$_d$. We then cool from 63 K down to 1.5 K while maintaining 1 GPa, we see no obvious change in the phase fraction of the T$_d$ phase, which is shown in figure \ref{fig3}(c). It should be noted that due to differences in the structure factor between the monoclinic reflections and the orthorhombic reflection, as well as monoclinic twinning, the orthorhombic reflection is significantly more intense than the monoclinic reflections which limits our ability to detect small phase fractions of monoclinic phase above our background level. While we see a reduction in the orthorhombic peak intensity by 0.4 GPa, we do not see intensity at the monoclinic position until 0.6 GPa, and we attribute this is to our detection limits. By tracking the angular position of the monoclinic reflections we can track the $\beta$ angle of the 1T' phase. We observe that $\beta$ increases with pressure, consistent with both our DFT calculations (Fig S2.) and with previously reported x-ray diffraction measurements \cite{qi2016superconductivity}. Importantly, as we observe intensity at the monoclinic peak positions, we see equal scattering intensity from both expected monoclinic twins in this zone indicating pressure homogeneity.  \par

Our study has uncovered a complex interplay between the crystal structure of this system and the underlying electronic ground state. Below 0.8 GPa, our transport measurements indicate partial volume fraction superconductivity and show a strong anomaly related to the T$_d$ to 1T' transition. The neutron diffraction measurements show that the phase fraction of the low-pressure T$_d$ phase also drops below 50\% at 0.8 GPa. In contrast, previous ac-susceptibility and $\mu$SR measurements indicate that within this pressure regime, full volume fraction superconductivity is achieved \cite{guguchia2017uSR}. The large phase coexistence region in both pressure and temperature suggests that the T$_d$ and 1T' phases are very close in energy. To address this interplay between pressure, structure, and superconductivity, we turn to first-principles calculations. In particular, we have calculated the pressure dependence of the stability of each phase, the reaction path between the measured structures, how the electron-phonon coupling changes between the Td and 1T' phases, and whether both structures would be expected to support superconductivity. The details of these calculations are given in the supporting information \cite{supplement2}. \par

Our total energy calculations indicate that (see Fig.S1) both phases are nearly degenerate and only separated by an energy barrier of 0.8 meV, in agreement with recent calculations \cite{kim2017origins} but in contrast to previous assumptions as to the origin of the large phase coexistence region between the T$_d$ and 1T' phases \cite{clarke1978,dawson1987electronic}. The centrosymmetric phase $1T'$ always has a slightly lower volume than the non-centrosymmetric $T_d$ phase with applied pressure and therefore at high pressure the enthalpy term prefers 1T' over $T_d$ as shown in Fig.S3 justifying the pressure driven suppression of T$_S$. We have also calculated full phonon dispersion curves for both phases at different pressures up to 10 GPa and did not find any phonon softening to explain this structural transition(see Fig.S4-S5) in contrast to \cite{duerloo2014structural}. Interestingly, the calculated free-energy when considering the full phonon dispersions at ambient pressure also prefers the 1T' phase over the $T_d$ phase at high temperatures as in the case of enthalpy. Hence, the observed phase transition is not soft-phonon driven but rather entropy driven. \par 
To better explain this non-intuitive result we offer the following explanation. Qualitatively, when viewed orthogonal to the orthorhombic b-c plane (as is shown in figure 1(b)), the Mo-Te zig-zag chains of atoms resemble opposed saw-teeth across the van der Waals bonding. If one were to slide these two sheets past each other along the orthorhombic b direction they would observe a periodic potential as the saw-teeth pass each other. As shown in figure S4(a), the inter-plane sliding mode along the long-axis is very anharmonic and features two shallow minima. In the lowest energy minimum, the MoTe2 planes (i.e. saw-teeth points) are more on top of each other and the curvature of one minima is slightly larger than the other. This results in slightly higher phonon energies and also gives a larger c-axis lattice parameter. When one of the planes slides a small amount and enters the minima along b, the teeth of the saw-like planes interlock, causing a c-axis contraction but lowering the energy required for a transverse motion along a, giving lower phonon energies and higher entropy. Hence, at high temperature the system prefers this interlocked configuration where the c-axis is shorter and inter-sliding phonons are lower in energy (i.e. higher entropy). This observation is consistent with the observed negative thermal expansion and the longer c-axis of the lower temperature Td phase.  When we cool the system, entropy is less important and the system prefers to be at the minimum energy configuration with the planes aligned on top of each other with a longer c-axis and orthorhombic symmetry, but higher phonon energies.\par

We have also calculated the electron-phonon coupling ($\lambda$) for both structures. Despite the strong apparent correlation between structure and superconductivity, the calculated coupling in both phases is very similar, indicating that the main contribution to superconductivity comes from within a single layer MoTe$_2$ unit. Indeed we found very similar $\lambda$ for both single layer MoTe$_2$ and bulk-like MoTe$_2$ (see SI. Section D)\cite{supplement2}. For both bulk-like phases and the single layer analogue, we find that all phonon modes contribute to $\lambda$. This phase independent and apparent isotropic and energy independent contribution to $\lambda$ suggests that there is some other contribution to superconductivity enhancement in MoTe$_2$ beyond the structural transition. The main difference between bulk-like and single layer MoTe$_2$ is found to be the pressure dependence of the $\lambda$. For the case of bulk MoTe$_2$ we did not find significant pressure dependence (Fig.S12) while for a single layer, T$_c$ is increased by an order of magnitude at 10 GPa pressure (Fig. S9)  as experimentally observed \cite{qi2016superconductivity}.  \par 

The single layer nature of $\lambda$ and the large structural phase coexistence region raises interesting questions about the nature and origins of superconductivity in MoTe$_2$. The previously observed full volume fraction superconductivity from ac-susceptibility and $\mu$SR in this coexistence region rules out the possibility that superconductivity is living only in isolated regions of the sample where single layers may be structurally decoupled. The interesting 2 gap model needed to explain the temperature dependence of $\lambda$$_{eff}$$^{-2}$ (where $\lambda$$_{eff}$ is the powder average effective magnetic penetration depth) in pressure dependent $\mu$SR could indicate that there is a different superconductivity living in the two phases \cite{guguchia2017uSR}. The nature of the interfaces between non-centrosymmetric and centrosymmetric regions of the sample in the mixed phase may further lead to novel physics and potentially different superconducting states between the two regions. These interfacial regions may even support novel band topologies given the broken symmetry at the interfaces and the possibility for a Weyl semimetal in proximity to a superconductor. 
The apparent single layer nature of the pressure dependence of $\lambda$ and the T$_c$ enhancement observed empirically hints that some kind of single layer decoupling happens with hydrostatic pressure which is surprising. This could be due in part to the expected large number of stacking faults for a Van der Waals bonded material, which have been demonstrated in MoTe$_2$\cite{yan2017investigation,manolikas1979electron}. This is not to say that we are creating new stacking faults with pressure, but rather that pressure appears to make the system more quasi-2D, which may be related to interactions and dynamics of pre-existing planar defects like stacking faults. Furthermore, while the $\mu$SR study did not consider this, if this pressure enhanced superconductivity is quasi-2D and there is a spin-triplet component to the pairing (allowed by symmetry) this may be a further route to topological superconductivity \cite{sato2017topological,sato2009,tanaka2009theory}. Further characterization of the nature of superconductivity in the system that considers the actual populations of the two structural phases and their interfaces is needed to explore these possibilities. 
\par
Since we now do not expect that enhanced superconductivity must live only in the centrosymmetric volume fraction of a crystal, we can ask whether there is a way to independently control crystal symmetry and superconductivity. Given the small energy difference between the two phases, one might expect that experimentally achievable strains might also influence the preferred crystal structure. Indeed, our calculations shown in figure 4(d) show that uniaxial strain can stabilize either the T$_d$ or 1T' phase depending on the crystallographic axis along which the strain is applied. \par 

In an effort to take advantage of the calculated uniaxial strain dependence of the ground state crystal structure, we have also performed structural measurements at the Oak Ridge National Laboratory High-Flux Isotope Reactor on the HB-3A four-circle diffractometer \cite{hb3a}. Here a clamp cell with a fluorinated pressure medium is used, similar to the cell described here \cite{aso2006neutron}. This fluorinated pressure medium has also been demonstrated to be non-hydrostatic above 0.8 GPa, leading to a non-hydrostatic pressure environment in our cell \cite{sidorov2005hydrostatic}. Figures 4 (a) and (b) illustrate the different cell geometries while (e) and (f) illustrate the difference stress environments within the cells. Here we have taken the standard notation where hydrostatic pressure corresponds to a stress tensor with equal and negative (compressive) diagonal components. With the clamping axis applying an larger uniaxial compressive stress along the monoclinic notation crystallographic b-axis, this is equivalent to negative strain along b shown in figure 4(d). At a clamp loading corresponding to 1.5 GPa with this media, we observe clear evidence of non-hydrostatic pressure in the form of strain broadening and detwinning of the monoclinic phase. We also observe a complete change in the ground state crystal structure. As shown in figure 4(g), at a nominal pressure of 1.5 GPa we lose all evidence of any monoclinic phase below 90 K (measured down to 5 K). Upon warming the previously defined monoclinic reflection starts to show up at 100 K and the phase transition is completed by 230 K. This is in contrast to the observed coexistence from our He cell measurements from figure 3 and figure 4 (c). Here we have clear evidence that the ground state crystal structure can be controlled though careful design of the mechanical stress environment, but also that structural determination is critical for interpretation of other measurements. Other groups have also noted the empirical importance of uniaxial strain in this system for magnetotransport properties and for T$_S$ at ambient pressure\cite{yang2017strain}. Our extracted single crystal lattice parameters and the change in a/b ratio under pressure loading in this clamp cell (shown in table S5) \cite{supplement3} are also consistent with a uniaxial stress geometry compared to the unloaded state . \par

The ability to stabilize the full volume fraction of the T$_d$ phase with non-hydrostatic pressure offers a simple route to a monophase non-centrosymmetric superconductor. Given our calculations of $\lambda$ in the two crystal structures, and given the full volume fraction superconductivity in polycrystalline samples from ac-susceptibility \cite{guguchia2017uSR}, we should expect that the enhancement in superconductivity is independent of the ground state crystal structure. One would expect no preferential phase selection in the polycrystalline system given the random orientation of grains with respect to possible non-hydrostatic pressure. Strain control of structure independent of superconductivity enhancement also explains the previous disagreements in pressure-temperature phase diagrams of T$_S$ defined by transport\cite{takahashi2017anticorrelation,qi2016superconductivity}. We can think of MoTe$_2$ as offering a system where pressure tunes superconductivity through shifting the single layer electronic DOS and possibly decoupling the layers while in-plane stresses (strains) can select between the centrosymmetric and non-centrosymmetric phases.  The huge stability window in both pressure and temperature of the mixed phase state offers a further unique opportunity for phase engineering in this system by tuning structural phase fractions.   \par

\section{\label{sec:level4}Conclusions}
Our results illustrate the possibility to independently control inversion symmetry breaking through structural manipulation in MoTe$_2$ as well as superconductivity in MoTe$_2$ using temperature, hydrostatic pressure, and the symmetry of non-hydrostatic components of pressure (uniaxial-like stress). This decoupling of the superconductivity from the structural transition explains previous disagreements between transport and magnetic measurement generated T-P phase diagrams \cite{qi2016superconductivity,guguchia2017uSR,takahashi2017anticorrelation}. We have shown the coexistence of the T$_d$ and 1T' phases at hydrostatic pressures and temperatures concurrent with full volume fraction superconductivity, which demonstrates that MoTe$_2$ can support topological superconductivity in certain regions of the sample, or in full sample volumes under non-hydrostatic pressure loading. The nature of this topological superconductivity can take multiple forms, whether through a proximity effect in the mixed phase region or through a full non-centrosymmetric bulk superconductivity in a Weyl semimetal. We anticipate that these results will help elucidate future interesting and useful transport properties in this material, and may offer a route towards a superconducting system with strain tunable Weyl Fermi arcs and non-trivial band topology.

\section*{Acknowledgements}

We acknowledge useful discussions with C. Brown.  Certain trade names and company products are identified in order to specify adequately the experimental procedure. In no case does such identification imply recommendation or endorsement by the National Institute of Standards and Technology, nor does it imply that the products are
necessarily the best for the purpose. This work utilized facilities supported in part by the National Science Foundation under Agreement No. DMR-0454672. This research used resources at the High Flux Isotope Reactor, a DOE Office of Science User Facility operated by the Oak Ridge National Laboratory. Research at the University of Maryland was supported by AFOSR through Grant No. FA9550-14-1-0332 and the Gordon and Betty Moore Foundation’s EPiQS Initiative through Grant No. GBMF4419.

\section*{Author contributions statement}
C. Heikes and N. Butch synthesized materials.  C. Heikes, W. Ratcliff II, Yan Wu, J. Leao, and Huibo Cao performed neutron scattering measurements.   Bulk property measurements were performed by I-Lin Liu, Tristin Metz,
Chris Eckberg, and N. Butch.  L. Hung and T. Yildirim performed DFT calculations. All authors contributed to the manuscript. 

\section*{Competing financial interests} The authors declare no competing financial interests.


\begin{figure}
\begin{center}
\includegraphics[width=\columnwidth]{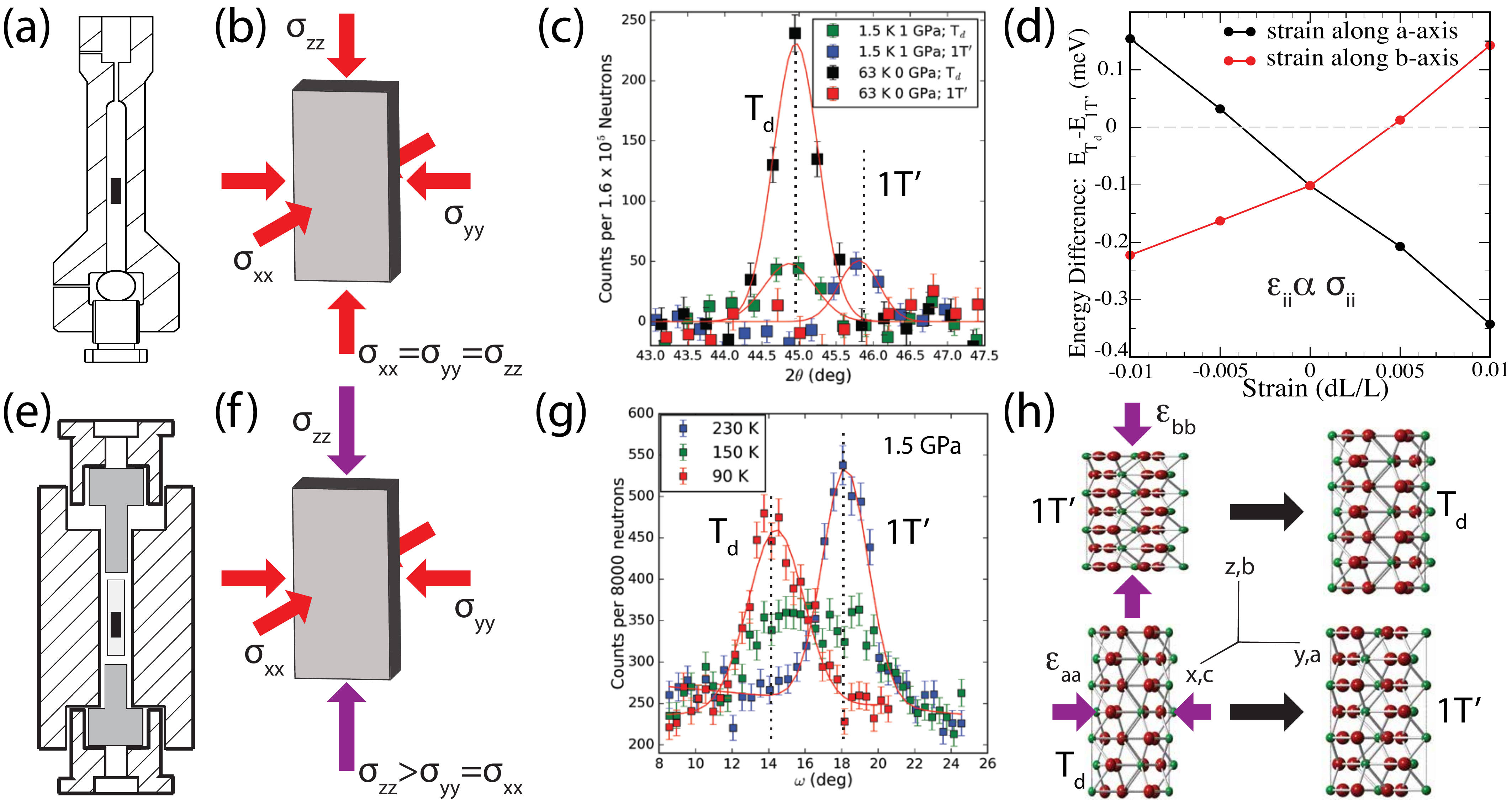}
\end{center}
\caption[Pressure cell dependent structure]{Effect of strain field on crystal structure.  (a). Schematic of of He gas cell used for hydrostatic pressure environment measurements. Gas is loaded externally into the cryostat (b). Stress environment for a plate-like sample in the He cell assuming no shear. Stress is given as components of the stress tensor, with the convention that hydrostatic pressure is negative stress (c.) Longitudinal scans along (021)$_O$ (labeled T$_d$) and (201)$_M$ (labeled 1T') positions at ambient pressure (63 K) and at 1 GPa (1.5 K) for He cell. No peak is observed from the 1T' phase at ambient pressure. (d). DFT calculations of the energy difference between the $T_d$ and $1T'$ phases as a function of strain along the a-axis (black) and b-axis (red) at ambient pressure. Compressive strain is negative by convention. (e) Diagram of the CuBe clamp cell used in the ORNL experiments. The sample is sealed in a capsule with fluoroinert pressure media and pistons uniaxially compress the capsule. (f). Stress environment for a plate-like sample in the CuBe cell assuming no shear. Uniaxial loading and non-hydrostatic pressure transduction leads to increased stress component along clamping direction. (g) Rocking scans at (021)$_O$ and (201)$_M$ peak positions through the phase transition in the CuBe cell at 1.5 GPa. Below 90 K, we see no evidence of the 1T' phase  (h).  Unixaial strain along b drives phase from 1T' to T$_d$, uniaxial strain along the a-axis drives a transition from the T$_d$ phase to the 1T' phase. Axes illustrate the correspondence between monoclinic a,b,c and x,y,z in the stress diagrams.}  \label{fig4}
\end{figure}

\pagebreak
\widetext
\begin{center}
\textbf{\large Supplemental Materials: Mechanical control of crystal symmetry and superconductivity in Weyl semimetal MoTe$_2$}
\end{center}

\title{Mechanical control of crystal symmetry and superconductivity in Weyl semimetal MoTe$_2$  \\
Supplemental Material}%

\author{
Colin Heikes,$^1$
I-Lin Liu,$^{1,2,4}$
Tristin Metz,$^2$
Chris Eckberg,$^2$
Paul Neves,$^2$,
Yan Wu,$^3$ 
Linda Hung,$^1$ 
Huibo Cao,$^3$ 
Juscelino Leao,$^{1}$,  
Johnpierre Paglione,$^2$
Taner Yildirim,$^1$ 
Nicholas Butch,$^{1,2}$ and William Ratcliff II$^1$}

\affiliation{\vspace{4pt}$^1$NIST center for Neutron Research, NIST, Gaithersburg, Maryland 20899, USA.
\\$^2$Center for Nanophysics and Advanced Materials, University of Maryland, College Park, Maryland 20742, USA.
\\$^3$Quantum Condensed Matter Division, Oak Ridge National Laboratory (ORNL), Oak Ridge, Tennessee 37831, USA.
\\$^4$Department of  Materials Science and Engineering, University of Maryland, College Park, MD 20742 USA.}

\maketitle
\date{\today}%

\beginsupplement
\section{Experimental Methods}

Powder samples were prepared using the standard solid state synthesis method using high purity Mo powder (99.999\% metals basis excluding W, Alpha Aesar), and Te shot(99.9999\%, Alpha Aesar). Single crystals were grown using the Te self flux method as described in \cite{rhodes2016impurity} using the same source metals as for the powder samples. \par Powder neutron measurements were performed on the BT-1 diffractometers at NIST using the Cu(311) monochromator option at 60' collimation and 1.540 $\AA$. Powder patterns were fit using the Reitveld method in GSAS-II to obtain lattice constants, atomic positions, and thermal parameters \cite{toby2013gsas}. NCNR single crystal measurements were performed on the BT-4 and SPINS triple axis spectrometers at NIST. Ambient pressure measurements on BT-4 were made with a 14.7 meV neutron beam with a collimation and filter setup of open-pg-40'-pg-sample-pg-40'-100' while pressure dependent measurements were made also at 14.7 meV with open-pg-80'-pg-s-pg-80'-100' collimation and filters where pg referes to pyrolytic graphite. The SPINS data were taken at 5 meV neutron energy with 80'-Be-s-80'-Be filter and collimator configuration, where Be refers to a liquid nitrogen cooled Be powder filter. Single crystal data was also taken with the HB-3A four-circle single crystal diffractometer at ORNL using a wavelength of 1.546 A from a bent perfect Si-220 monochromator \cite{hb3a}. Refined crystal structure data is shown in supplementary tables S2, S3, and S4.   \par

Pressure dependent measurements on BT-4 were performed over a range from 0 to 1 GPa using a steel measurement cell and He as the pressure media as described in \cite{PhysRevBpressureneutron}. For these measurements, He is used to reduces possible pressure inhomogeneity compared to fluorocarbon-based techniques as has been demonstrated for CrAs or other pressure sensitive correlated electron systems \cite{PhysRevBpressureneutron,butch2009hydrostaticity}. The integrated peak intensities for phase fraction determination were obtained by fitting Gaussian peak profiles and a Lorentzian background to the raw scattering data to account for sample environment related background scattering. The HB-3A measurements investigated the crystal up to 1.5 Gpa pressure with a CuBe clamp pressure cell using Fluorinert 70:77=1:1 as the pressure transmission medium.  \par A non-magnetic piston-cylinder pressure cell was used for transport measurements under pressure up to 1.6 GPa, choosing a 1 : 1 ratio of n-Pentane to 1-methyl-3-butanol as the pressure medium and the superconducting temperature of lead as pressure gauge at base temperature. For transport measurements, we prepared a 110 $\mu m$ thick sample of MoTe$_{2}$ curing contacts with silver epoxy. Measurements were performed in magnetic fields up to 14 T and down to 1.8 K in Physical Property Measurement System (PPMS). For superconducting temperature below 1.8 K, resistivity measurements down to 25 mK in a dilution refrigerator were taken using a Lakeshore LS370 AC resistance bridge. The resistivity values were taken by the average of 60 stable and successive measurements. \par
\section{First-Principles Calculations}

\subsection{Method}
The first-principles  total energy, structure optimization under pressure and enthalpy calculations are performed by the Vienna {\it ab initio} simulation package 
VASP\cite{vasp1,vasp2}, 
which is based on density-functional theory (DFT),
using a plane wave basis set and the all-electron projected augmented wave 
(PAW) potentials\cite{paw1,paw2}.
The exchange-correlation interactions are described by the generalized gradient approximation (GGA)
with PBE type functional. The weak inter-layer van der Waals (vdW) interactions are included by 
optB86b functional\cite{optB86b}.  The Brillouin-zone
integration are performed using Monkhorst–Pack grids of special points with
$16\times8\times4$ for total energy and structure optimizations and $32\times16\times8$ with optimized-tetrahedra method for electronic density of states calculations.  The kinetic energy cut-off of 500 eV is
used in all calculations. For particular cases, the spin-orbit (so) interactions are included in
the calculations but the effect of SO-coupling is found to be minimal.

For the phonon dispersion curves and the electron-phonon coupling constants calculations,
we used  Quantum Espresso,\cite{Giannozzi2009} PAW pseudopotentials, the Perdew-Burke-Ernzerhof exchange-correlation functional,\cite{Perdew1996} $16\times8\times4$ $k$-point sampling, and 0.02 Ry Methfessel-Paxton smearing with wavefunction and charge density cut-off energies of 100 Ry and
800 Ry, respectively. The electron-phonon coupling constant is calculated in a denser k-grid of 
$36\times18\times8$ and q-grid of $6\times3\times1$. The vdw and SO interactions are also included. We used grimme-d2\cite{grimme} vdW correction with parameter $london-s6=0.6$.

\subsection{Energetics of $1T'-$ and $T_d$-phases}

In this section, we compare the energetics of the centrosymmetric $1T'$ and
non-centrosymmetric $T_d$ phases and try to explain the origin of the phase
transition between these two phases within the pressure-temperature plane.

\begin{figure}
\includegraphics[scale=0.6]{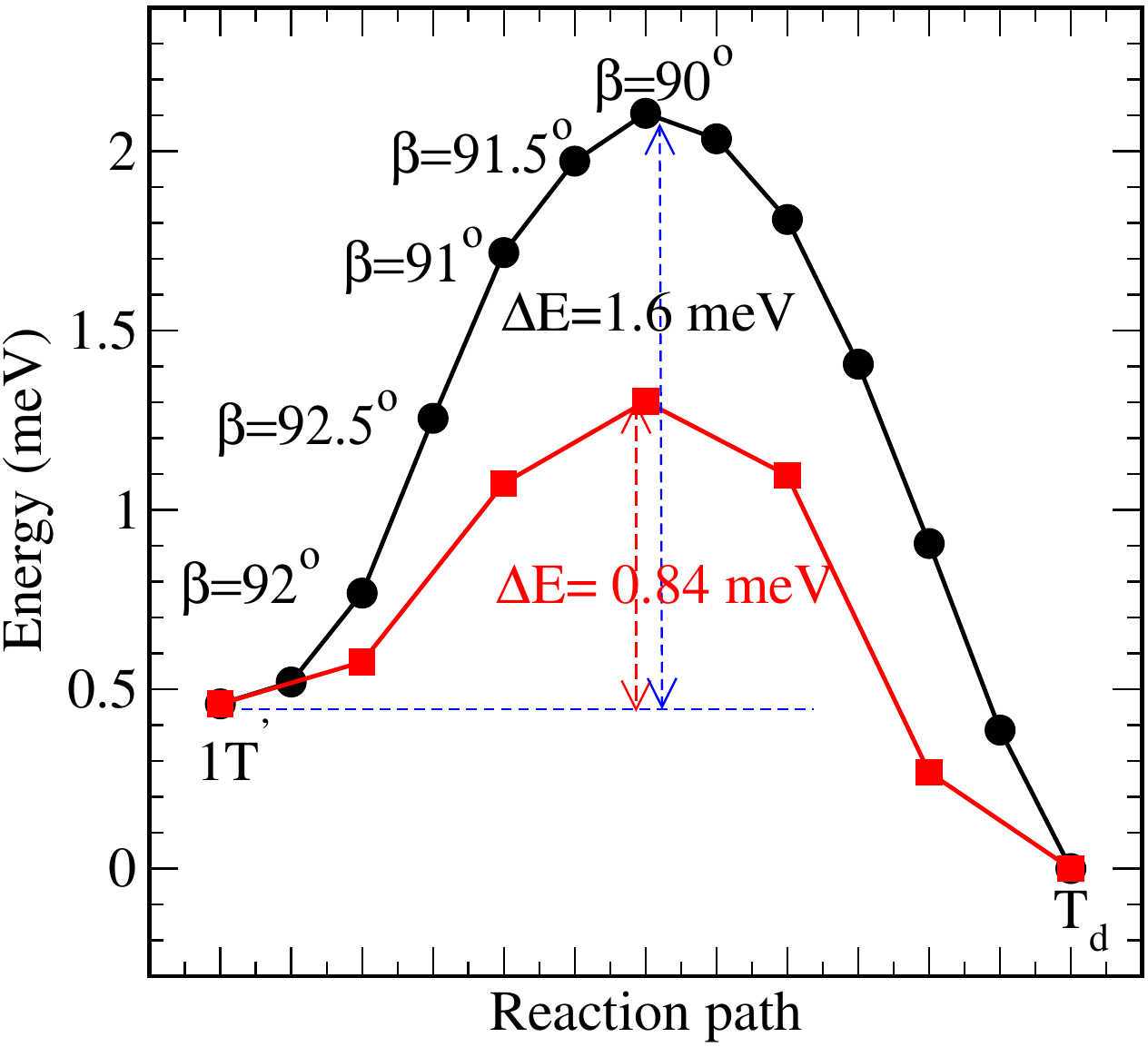}
\caption{\label{Td-T1p-NEB}  The energy barrier between two
phases $1T'$ and $T_d$ phases. The black curve is a simple path
as the angle $\beta$ is varied from its optimized value to 90$^o$, followed
by atomic optimization. The red curve is the optimized path by
nutget elastic band calculations. 
}
\end{figure}

Even though the two phases $1T'$ and $T_d$ look quite different with and
without inversion symmetry, we found that both phases are almost degenerate in
energy. Both VASP and Quantum Espresso give ground state energies that are
almost equal to within 0.5 meV for both phases. Figure~\ref{Td-T1p-NEB} shows
the energy difference between these two phases and the energy barrier between them
as obtained from VASP.  With no external pressure, we found that the
non-centrosymmetric phase $T_d$ is about 0.35 meV lower in energy than
the centrosymmetric $1T'$ phase. In order to get an idea about the energy
barrier between these two phases, we carried out nudget elastic  band NEB reaction
path calculations based on simple initial positions that are a linear superposition
of the two phases. The black curve in Figure~\ref{Td-T1p-NEB} shows the
energy barrier when the angle $\beta$ is reduced from the optimized value of
$93.3^o$ to $90.0^o$, which is about 1.6 meV. Once the angle is $90^o$, the
atomic positions in the $1T'$ phase go to the atomic position of the $T_d$ phase
lowering the total energy without any energy barrier. The red curve is
the result of the NEB calculations, indicating that the actual barrier between
these two phases is even less than 1.6 meV and around 0.8 meV.

\begin{figure}
\includegraphics[scale=0.5]{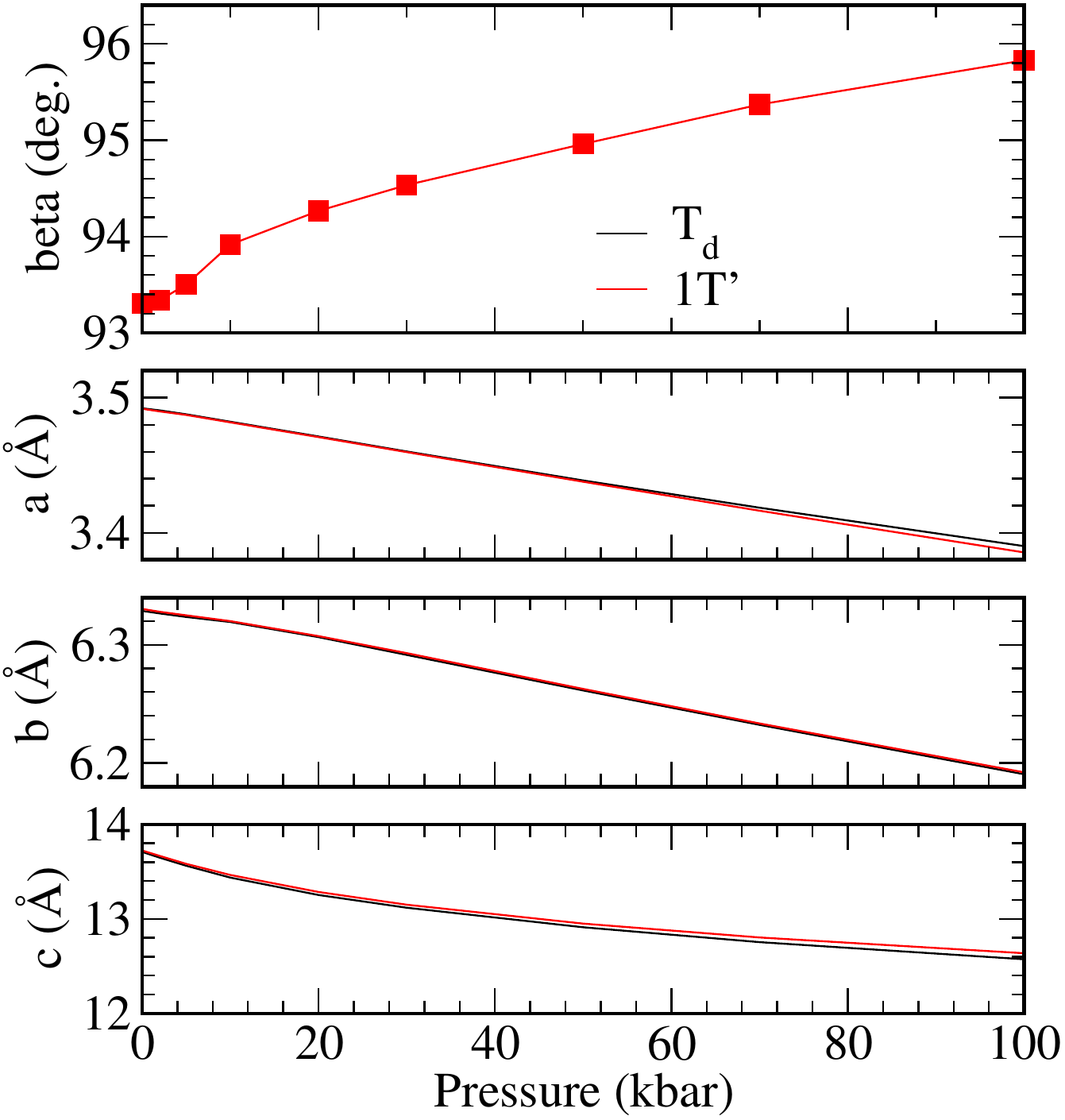}
\caption{\label{abcbeta_P}  Structural parameters with applied
pressure for both centrosymmetric $1T'$ (red) and non-centrosymmetric
$T_d$ phases (black).  We note that 1 kbar = 0.1 GPa. 
}
\end{figure}
 
Since the ground state energies of these two phases are so close, it is interesting
to see what stabilizes one phase over the other with applied pressure and temperature.
We performed full structure optimization with pressure for both phases and the
results are summarized in Figure~\ref{abcbeta_P}. The
cell angle $\beta$ is found to increase with applied pressure up to $96^o$,
which is consistent with our experimental observations. 
The $1T'$ phase always has a slightly lower volume than the $T_d$ phase, which suggests that at high pressure, due to
enthalpy term $P\times Volume$, the $1T'$ phase may become the ground state.

\begin{figure}
\includegraphics[scale=0.5]{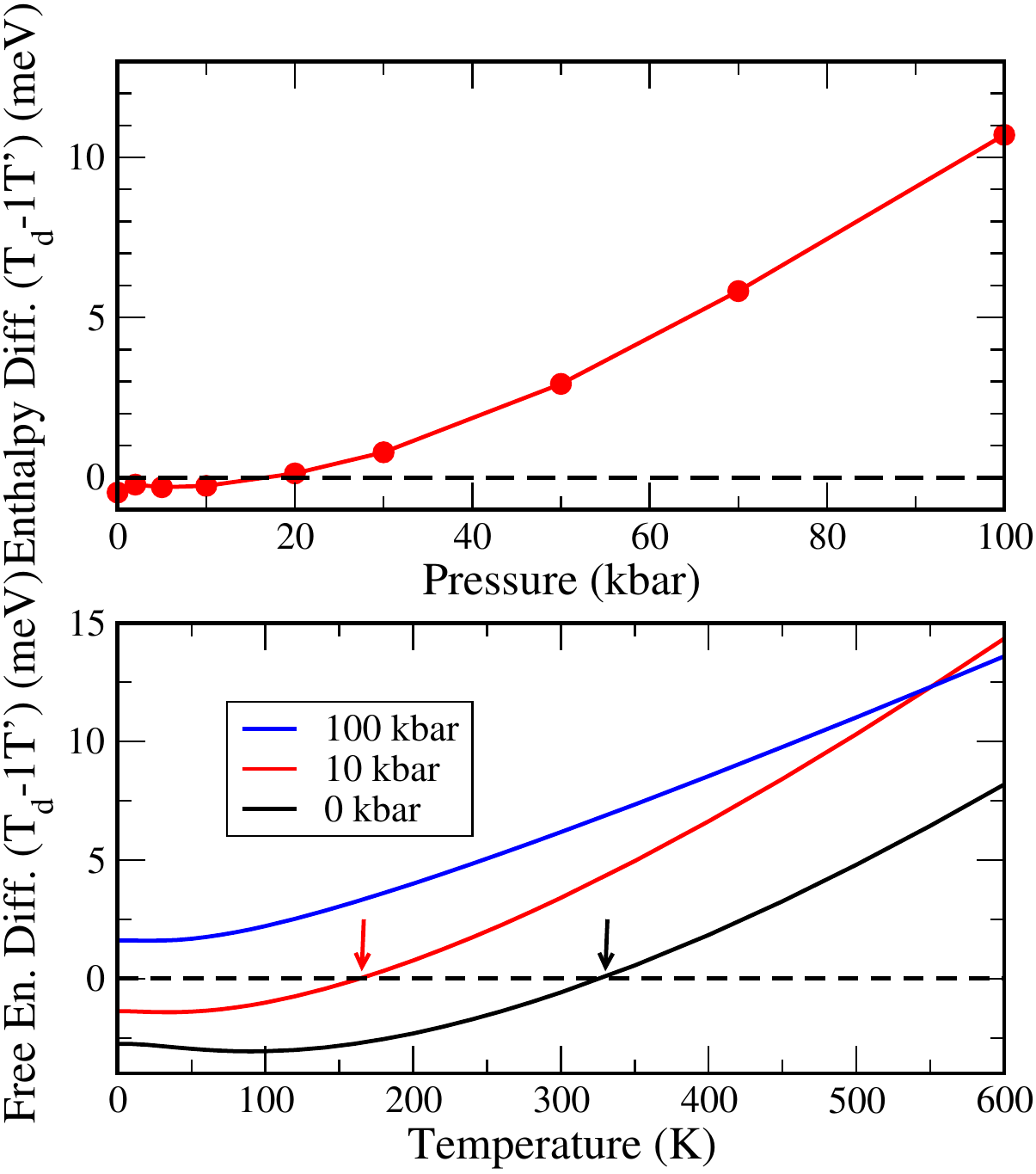}
\caption{\label{enthalpy} Top: The enthalpy difference (i.e.
$E + P \times Vol$) between $T_d$ and $1T'$ phases,
indicating a phase transition occurs near 20 kbar, above which the
$1T'$ phase is the ground state. 
Bottom: The free-energy difference between the two phases indicating
that the entropy term prefers the $1T'$ phase at high temperatures.
}
\end{figure}

Figure~\ref{enthalpy} shows the enthalpy difference between two phases with
applied pressure. The energy difference is still small up to pressures
of 20 kbar.  Only above 20 kbar does it increases rapidly, 
stabilizing the $1T'$ phase over $T_d$ above 20 kbar. Hence, it seems to be
possible to stabilize the $1T'$ phase with applied pressure. One may wonder
what is the mechanism for the phase transition at no external
pressure? In order to shed some light on this, we have calculated the full phonon
spectrum at a given pressure, from which we estimate the free energy of the system as a function of temperature. The bottom panel in Figure~\ref{enthalpy} shows the
difference in free energy between these two phases. Similar to the enthalpy term, the
free energy difference also prefers the $1T'$ phase at high temperature and high pressures.
At zero pressure, we see a sign change (i.e. phase transition) near 340 K which is
reduced significantly down to 160 K with 1 GPa pressure and then the free energy
difference is always positive for pressures larger than 10 GPa (i.e.the $1T'$ phase is the
ground state at pressures above 10 GPA at all temperatures). This trend seems to
be in perfect agreement with what we see experimentally.

\begin{figure}
\includegraphics[scale=0.3]{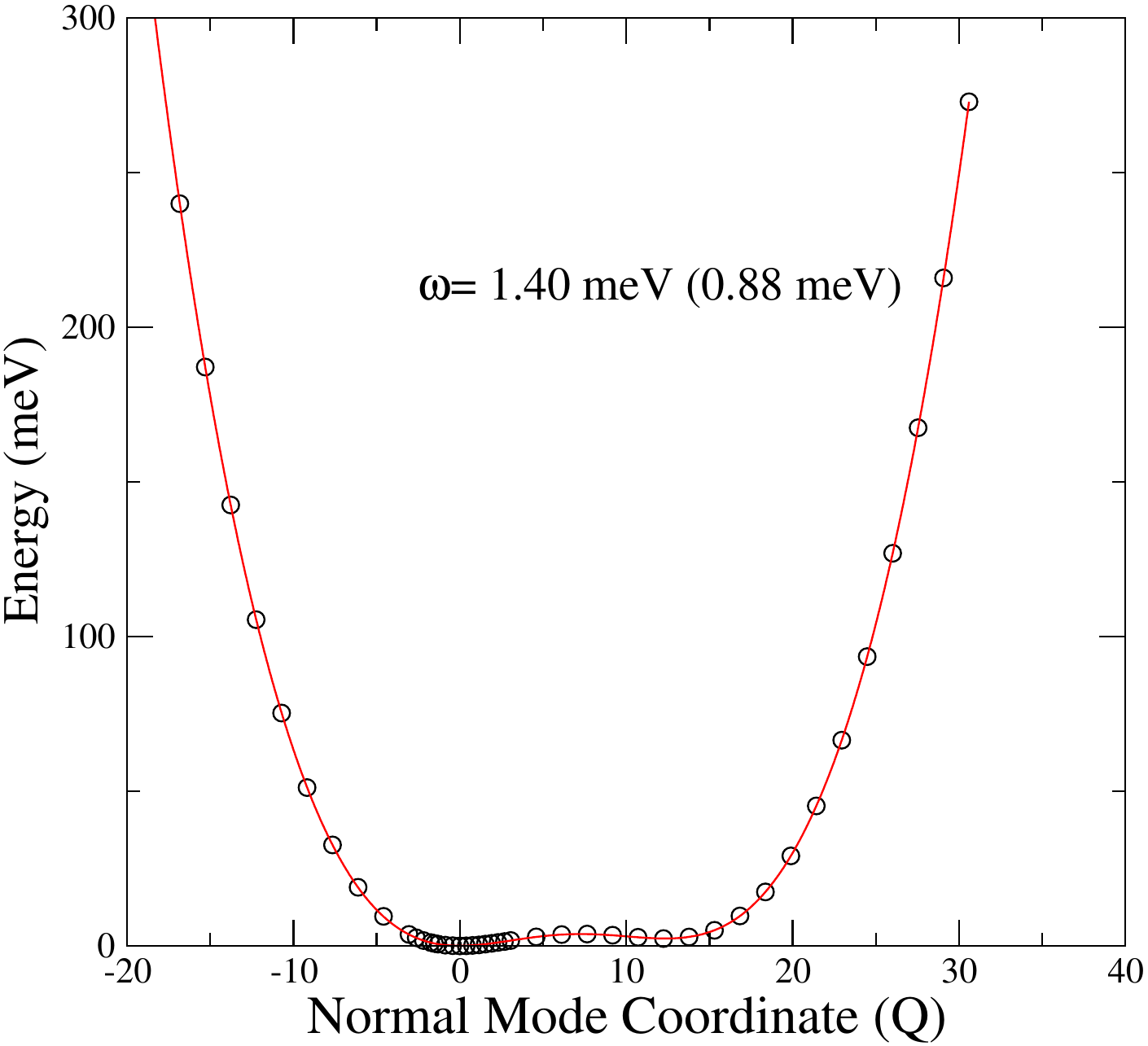}
\includegraphics[scale=0.3]{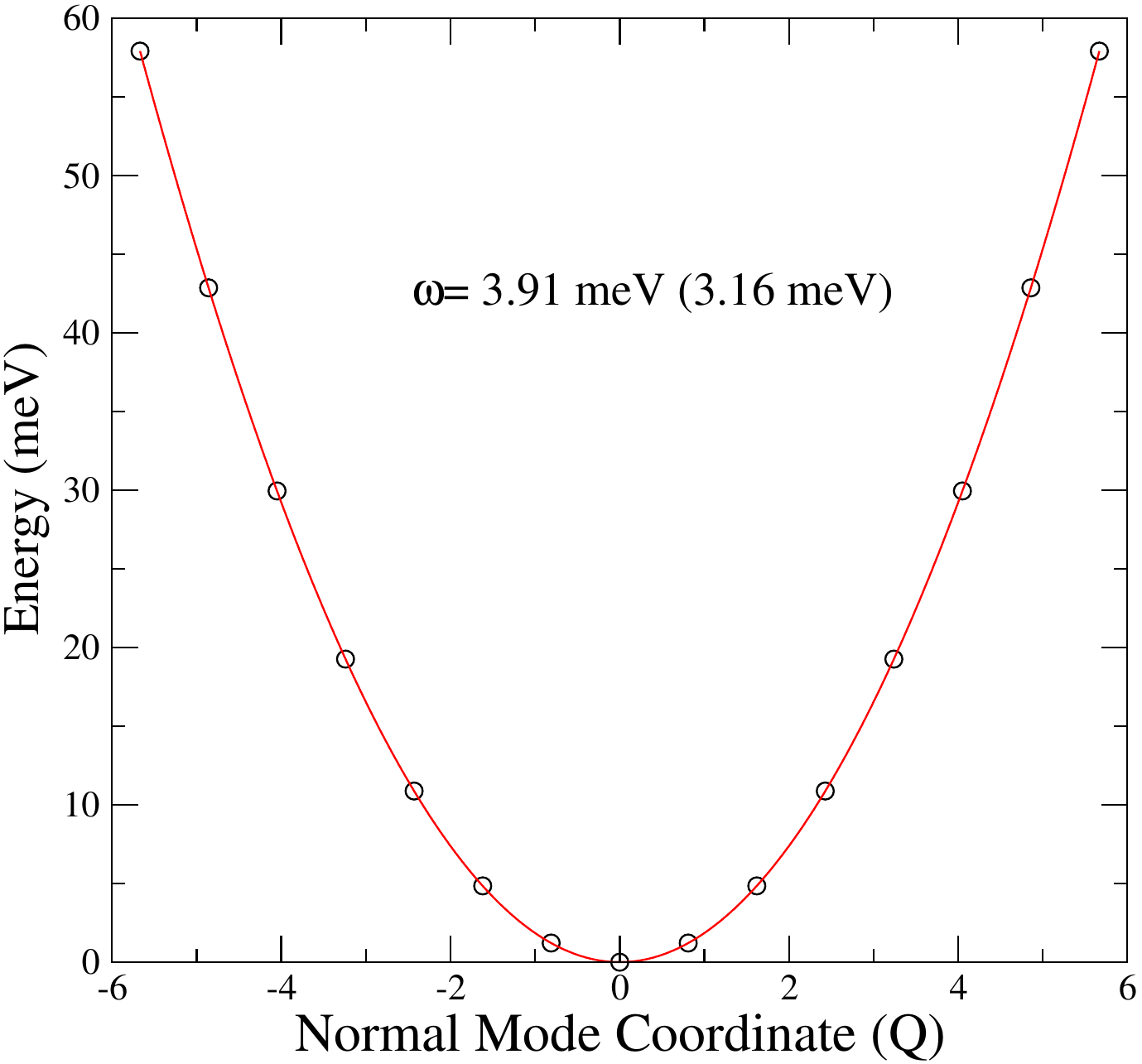}
\includegraphics[scale=0.3]{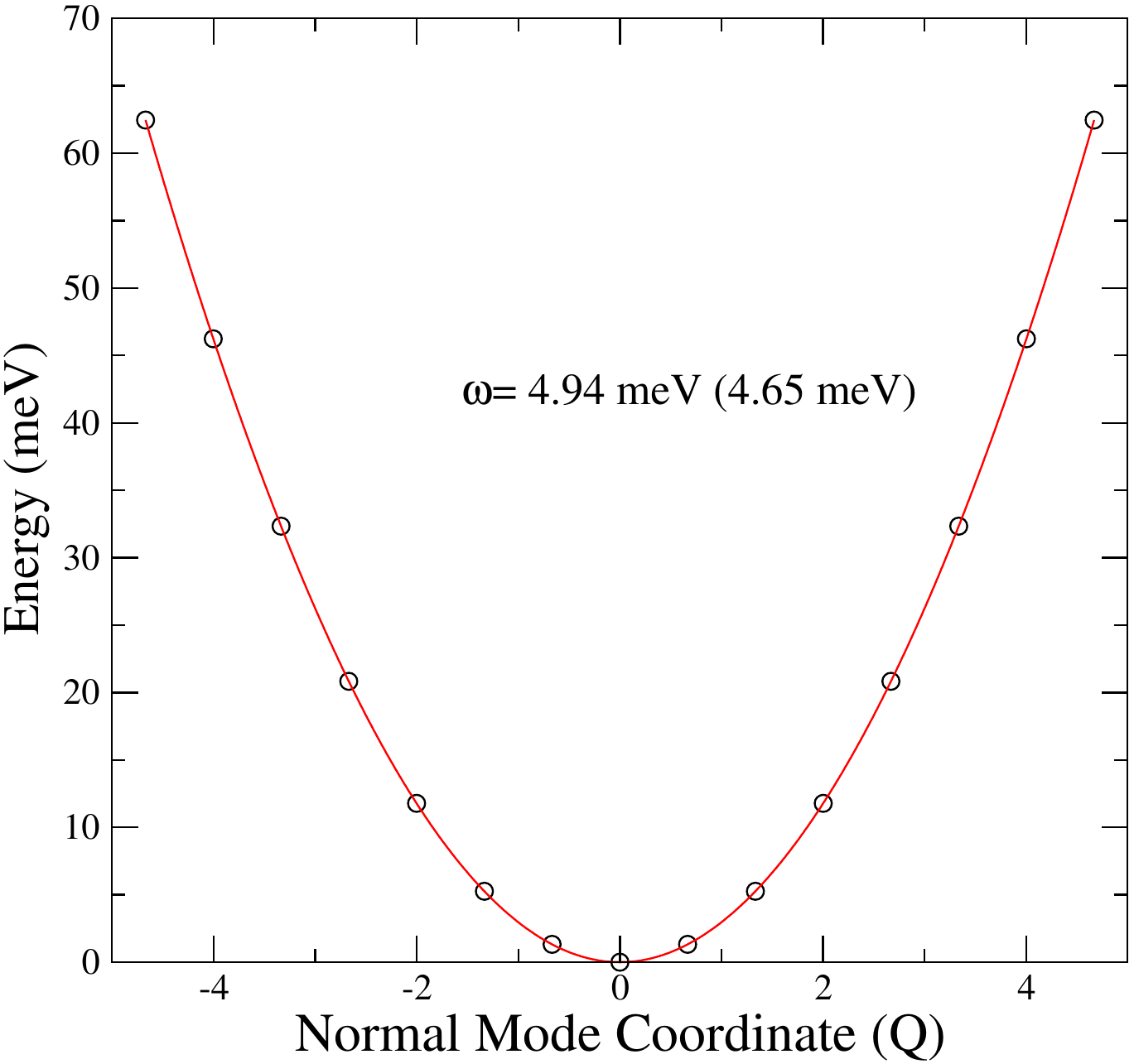}
\caption{\label{frzmodes} The total energy versus
the three lowest energy optical modes in $T_d$ phase.
The mode energies from solving the 1D Schrodinger
equation are also given. The numbers in parenthesis
are from the harmonic phonon approximation.
}
\end{figure}

\begin{figure}
\includegraphics[scale=0.275]{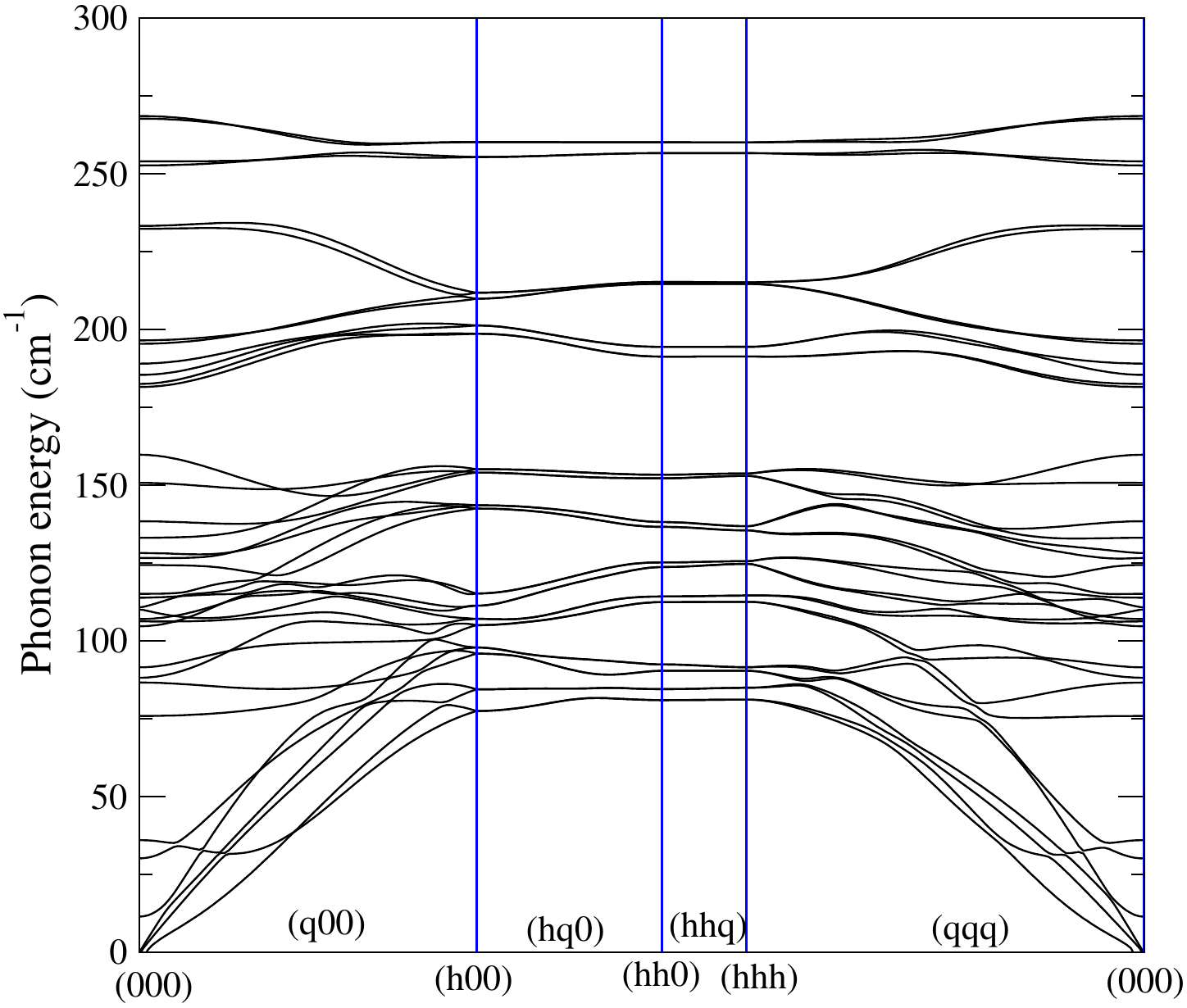}
\includegraphics[scale=0.275]{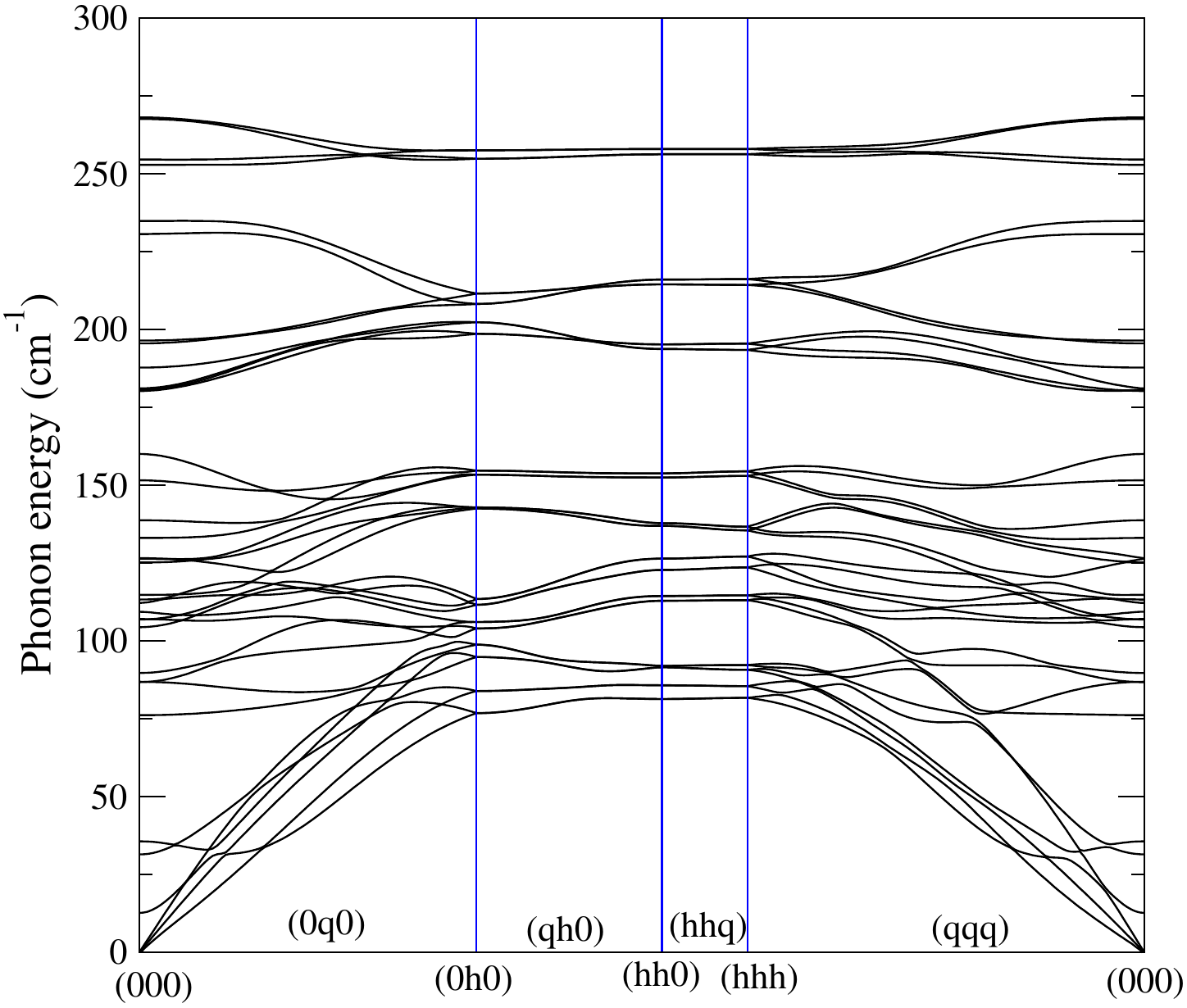}
\includegraphics[scale=0.275]{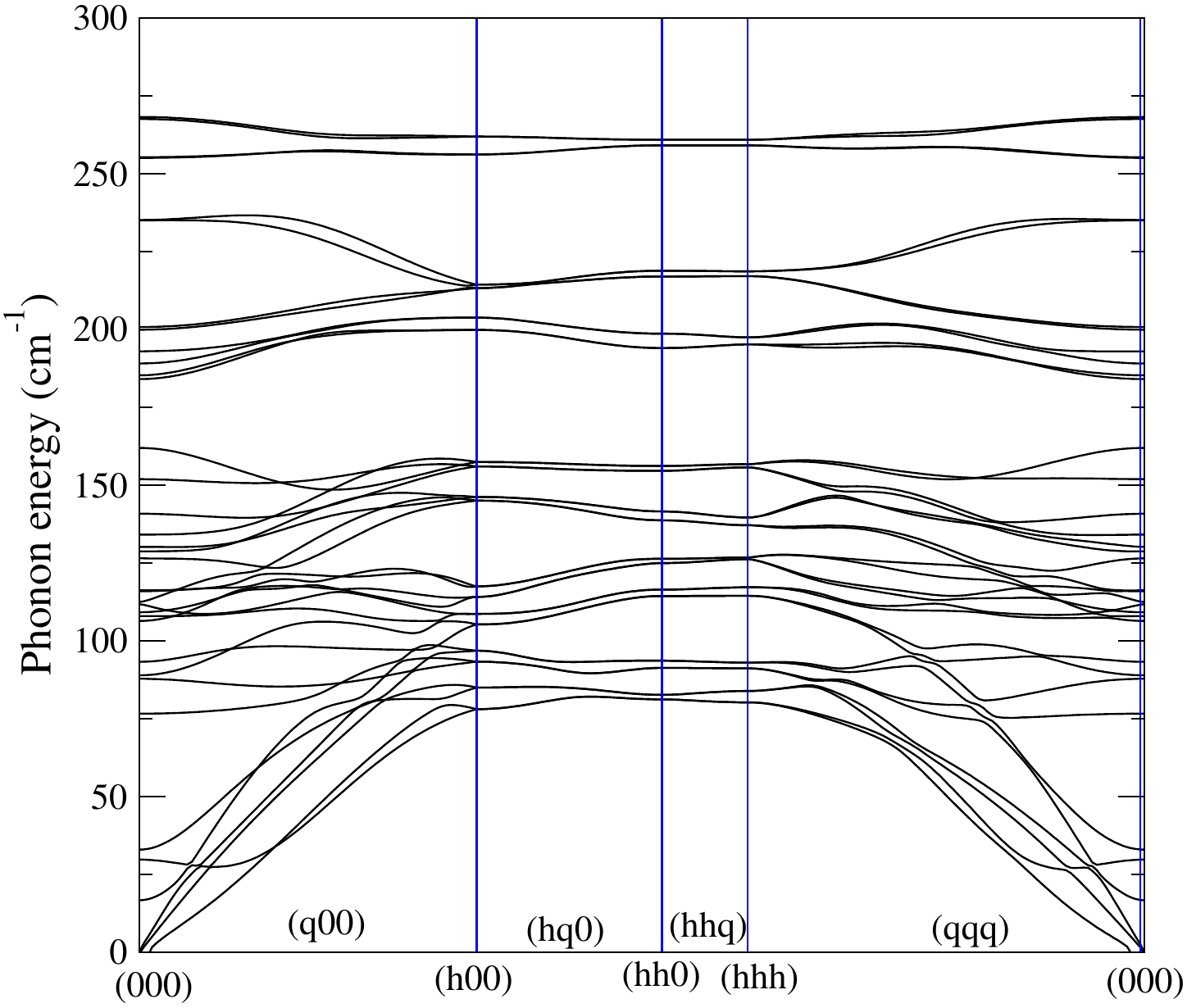}
\includegraphics[scale=0.275]{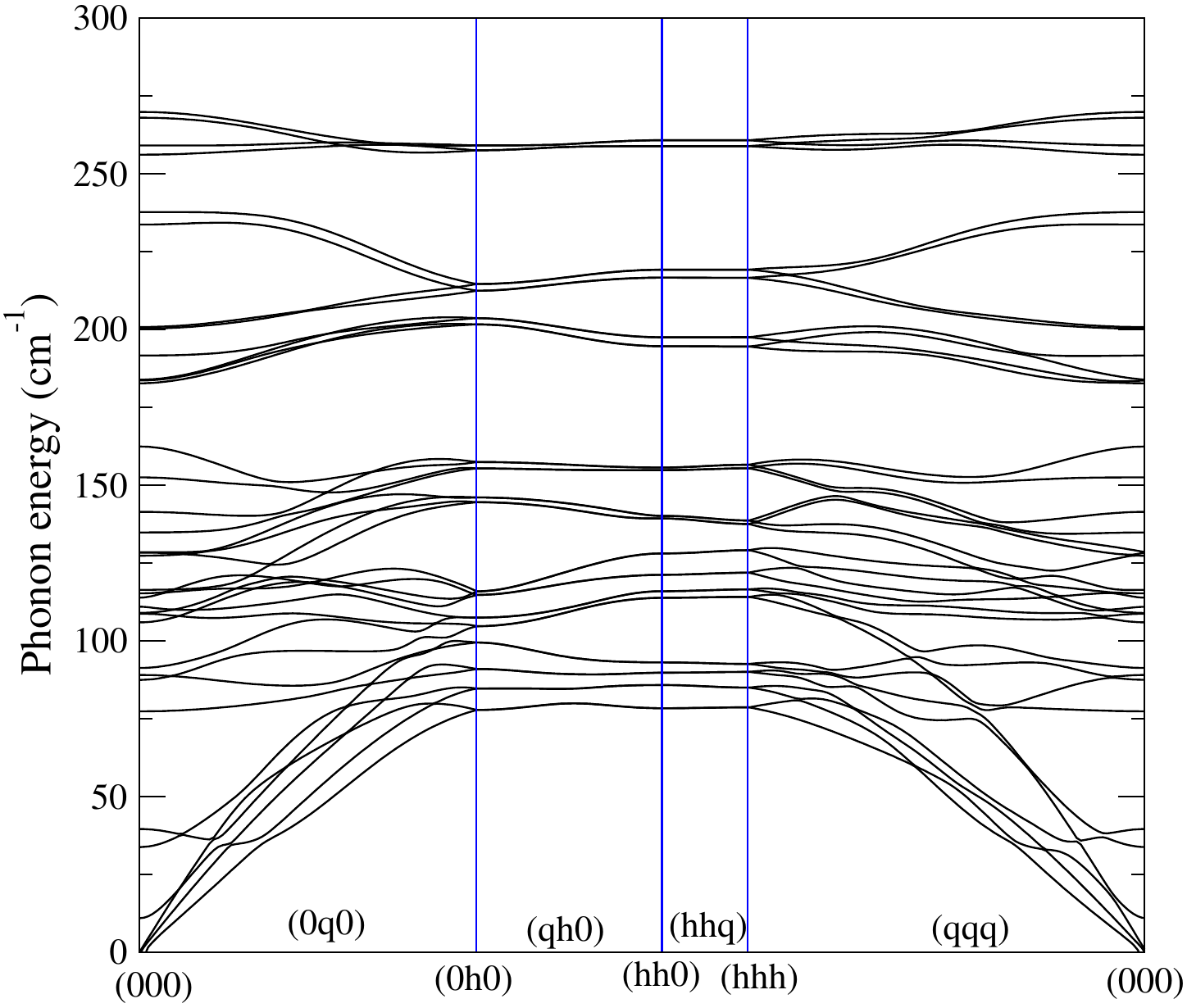}
\includegraphics[scale=0.275]{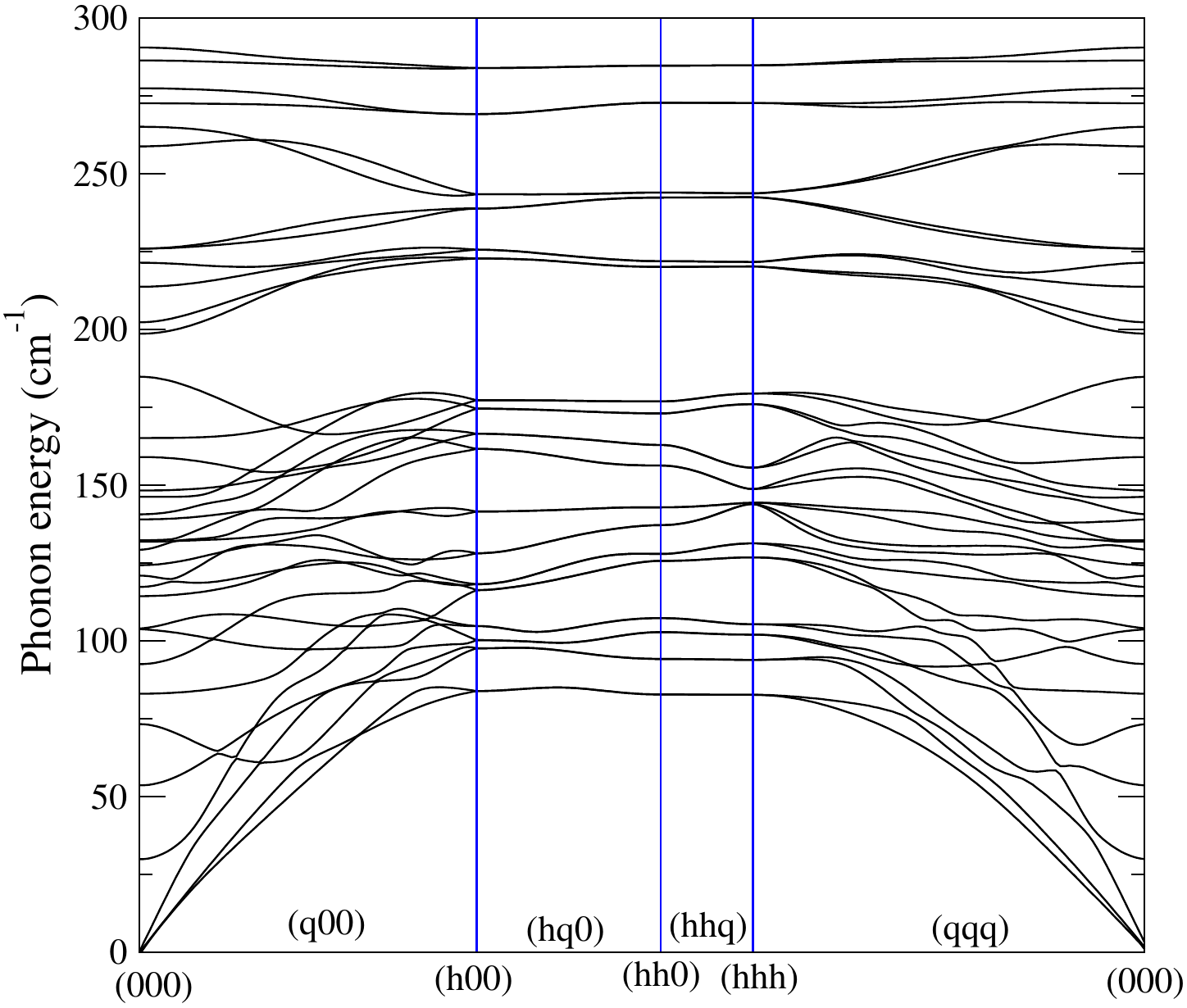}
\includegraphics[scale=0.275]{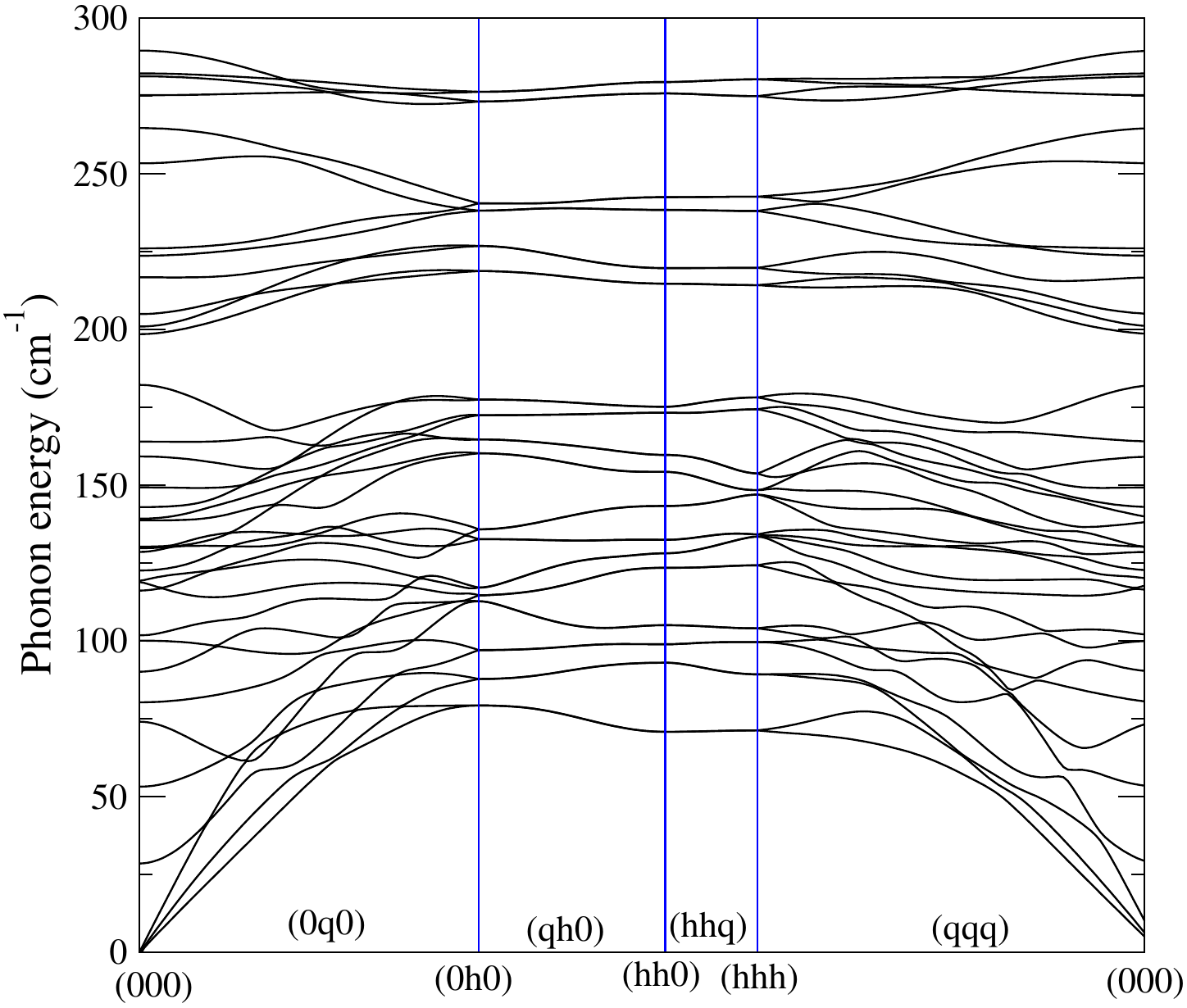}
\caption{\label{phdisp} Left: Phonon dispersion curves
for $T_d$ phase at P=0 (top),
P=10 kbar (middle) and P=100 kbar (bottom).
Right: Same but for $1T'$ phase.  No sign
of phonon softening with pressure in both phases.
}
\end{figure}
 
As an alternative mechanism for the phase transition from $1T'$ to 
$T_d$ phase, we also investigate the possibility of a soft phonon mode,
such as interplane sliding phonons of MoTe$_2$ planes as this type of
phonon would have a very low energy and it is also related to the cell angle
$\beta$, which defines how the layers stack.
Figure~\ref{frzmodes} shows the three lowest energy optical modes in
$T_d$ phase. Interestingly, the mode involves the inter-plane phonon oscillation
along the b-axis (i.e. direction related to cell angle $\beta$), is very
anharmonic, and has two slight dips in the potential. We repeated these frozen
phonon energies with applied pressure and did not see any negative modes with
applied pressure. The full phonon dispersion curves with pressures up to 100 kbar
are also calculated for both phases (shown in Figures~\ref{phdisp}) and no
sign of negative modes is observed. With applied pressure the lowest
energy optical modes shift up in energy and becomes more harmonic with 
applied pressure. Hence, it is very tempting to conclude that
the observed phase transition from $1T'$ to $T_d$ phase with temperature is
entropy driven, which is rather interesting. We are currently carrying out more
detailed and accurate calculations to include the effects such as temperature
dependence of the phonon energies to see if we can explain the phase transition
along with the observed negative thermal expansion for the c-axis. Based on our
results, it seems that the inter-planer coupling of the layers is quite crucial
not only to explain the negative thermal expansion but also the stabilize the
non-centrosymmetric ground state $T_d$ phase as we did not see a non-centrosymmetric
phase for an isolated single layer MoTe$_2$.

\subsection{Ground State Selection by Strain}

\begin{figure}
\includegraphics[scale=0.6]{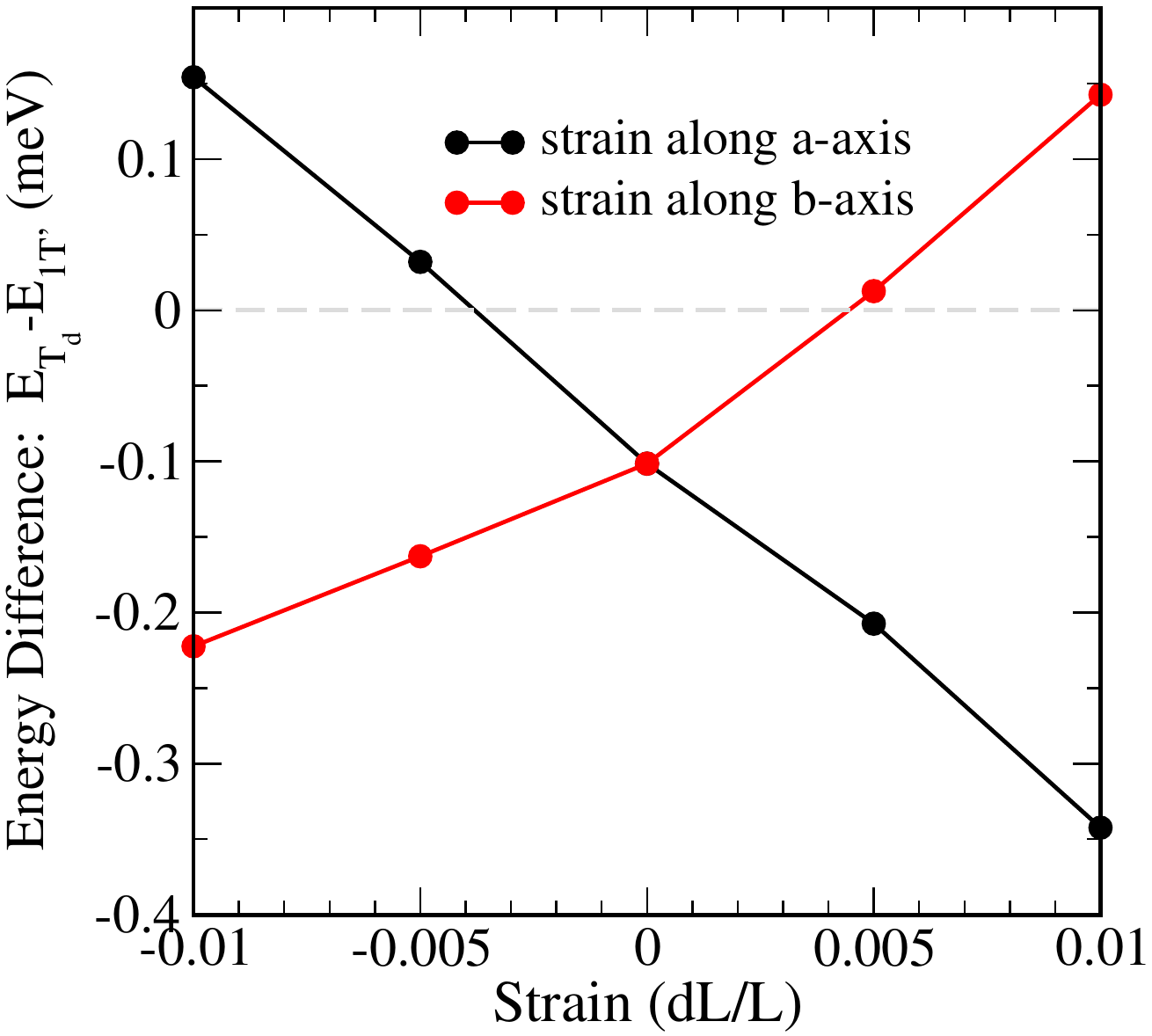}
\caption{\label{strain1} The energy difference between
$T_d$ and $1T'$ phases as a function of strain along
the a-axis (black) and b-axis (red) at 0 kbar (i.e. all other axes and internal coordinates are optimized). A strain of $\epsilon_a ~ -0.01 $ (i.e. compression along a-axis) is enough to change the sign of the energy difference from negative to positive, indicating the stabilization of $1T'$-phase at this strain. Similarly compression along b-axis stabilize the  non-centrosymmetric phase $T_d$.
}
\end{figure}

In previous section, we showed that by applying hydrostatic pressure, we can change
the energetics between $T_d$ and $1T'$ phases and make the centrosymmetric phase $1T'$ ground state above $~20 $ kbar pressure. Here we discuss the effect of anisotropic pressure by applying strain along one of the a- and b-axes. We define the strain as $\epsilon_a = \Delta a/a$; Hence the negative strain corresponds to compression while positive strain corresponds to elongation of the axis.
While we apply strains along one of the a- and b-axis, we let the other lattice parameters/angles and
the internal atomic positions fully relaxed. In this way, the change in the b-axis and c-axis in response to a-axis strain can be calculated and from the ratio we obtain the Poisson ratios. The total energy difference between the $T_d$ and $1T'$ phases
is shown in Figure~\ref{strain1} as a function of strain along a-axis (black) and b-axis (red).

The net effect of strain along a- or b-axis seems to be the opposite for a given compression or elongation, providing us a mechanism to switch the ground state.
When we apply compression along the b-axis, the
cell-angle $\beta$ decreases whereas if the b-axis is elongated, then the $\beta-$angle is also increased. It seems that the larger the $\beta$-angle, the more stable the $1T'$-phase. Hence when we apply pressure along the a-axis (which effectively increases the b-axis and in turn increases the $\beta-$angle) we stabilize the centro-symmetric $1T'$-phase. On the other hand, when we compress the system along b-axis, the non-centrosymmetric phase is stabilized. Hence by controlling the strain direction, one can tune the topology of the ground state.

\begin{figure}
\includegraphics[scale=0.4]{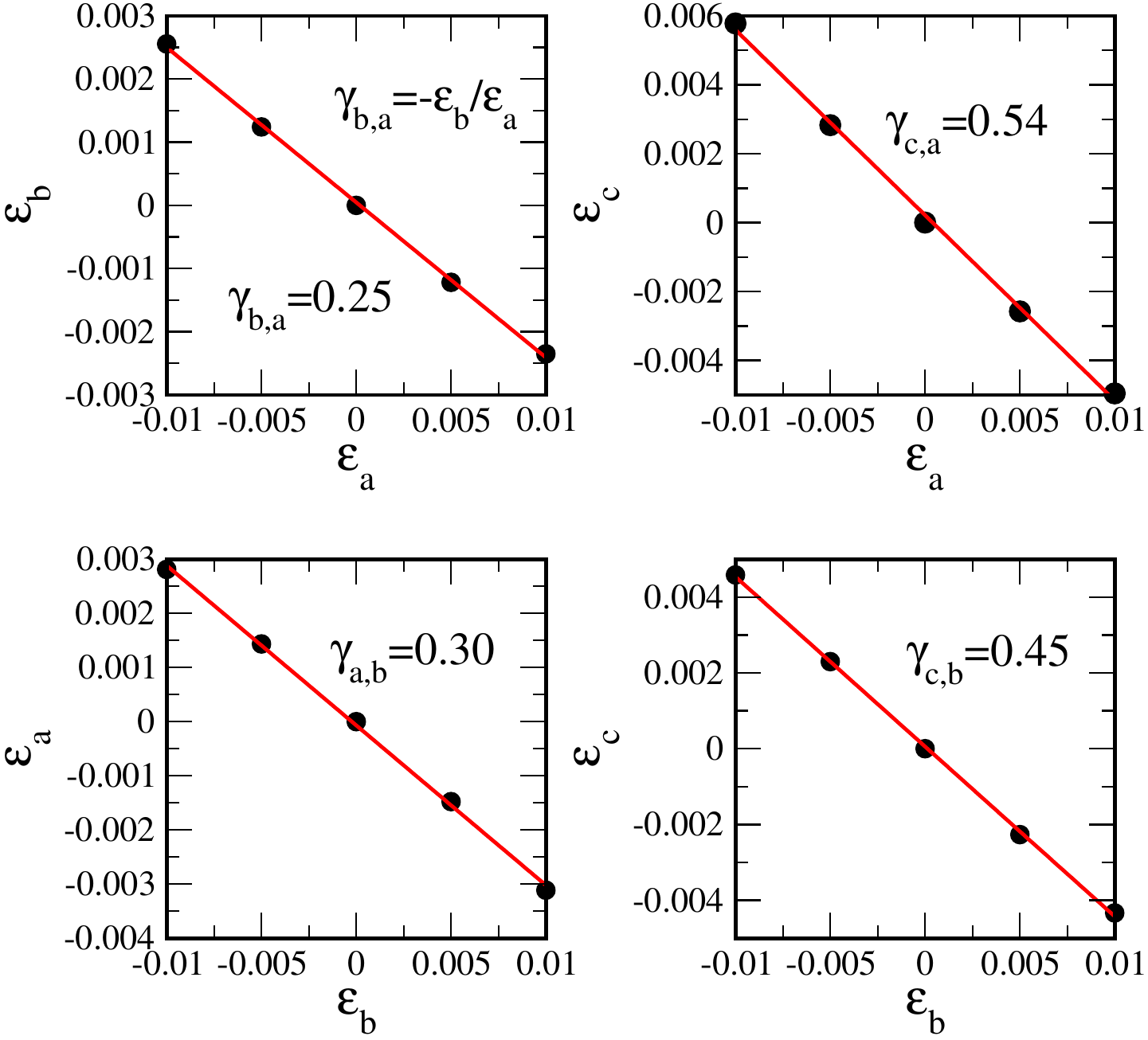}
\caption{\label{strain2} The response of the b- and c-axis
to strain along a-axis (top) at zero pressure for $T_d$-phase. 
Bottom panels shows the response of $a-$ and $c-axis$ 
to strain along b-axis.
The Poisson ratios are also indicated.
}
\end{figure}

\begin{figure}
\includegraphics[scale=0.4]{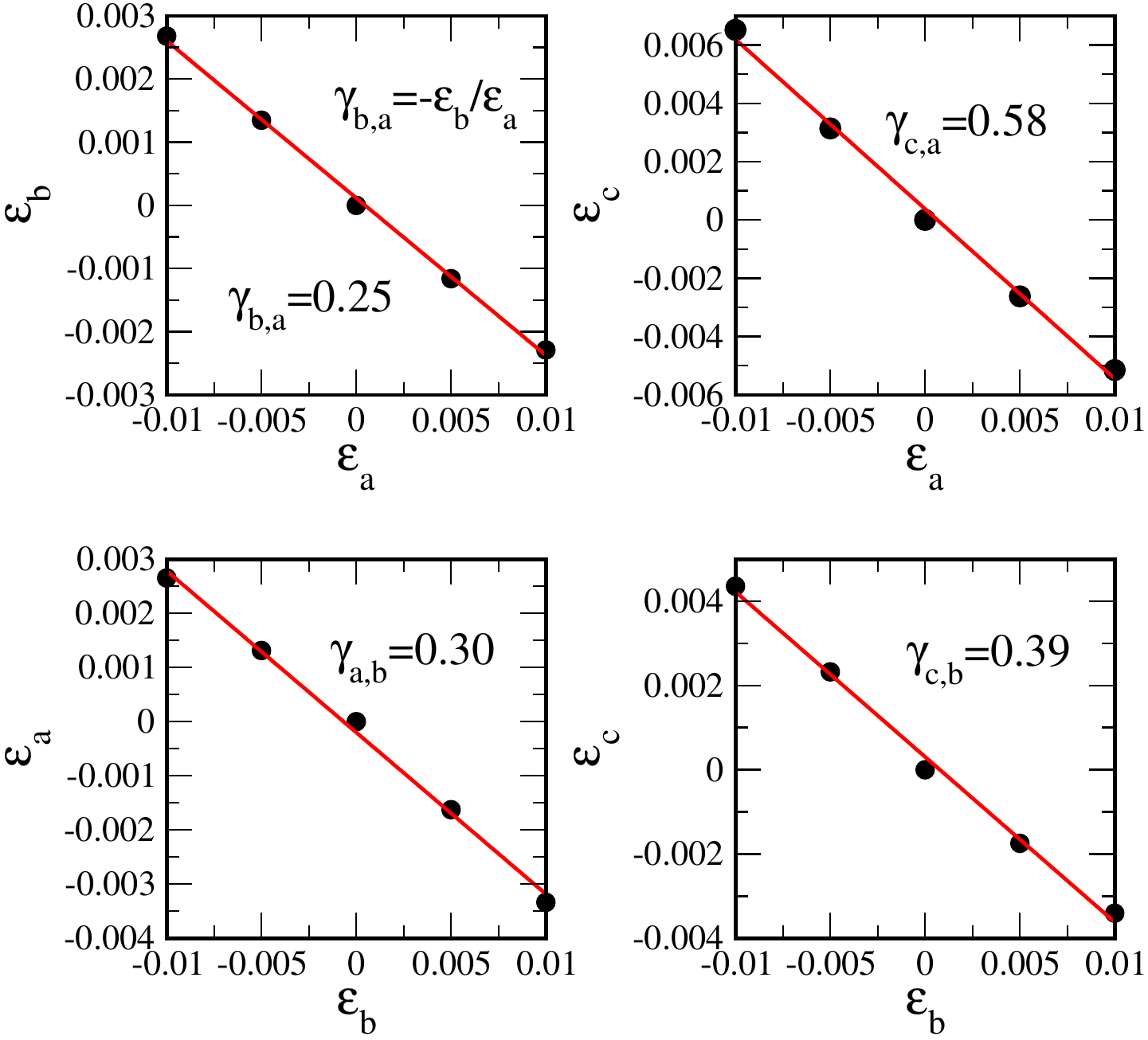}
\caption{\label{strain3} The response of the b- and c-axis
to the strain along a-axis (top) at 0 kbar in $1T'$ phase. 
Bottom panels shows the response of $a-$ and $c-axis$ 
to the strain along b-axis.
The poisson ratios are also indicated.
}
\end{figure}

From our constrained cell optimization under strain, we obtain the response of the
b- and c-axis to the strain along a-axis as shown in Figure~\ref{strain2}-ref{stain3}.
Compression along the a-axis causes elongation along the other two axes, yielding
positive Poisson ratios for MoTe$_2$.  Figures~\ref{strain2}-\ref{strain3} show
that the in-plane Poisson ratios for the $T_d$ and $1T'$-phases are the same. This is expected as the
response of a-axis to a change in b-axis and vice versa is mainly controlled by the in-plane interactions within a single MoTe$_2$ plane. The effect of vdW interactions between the planes seems to have a very small effect on the in-plane Poisson ratios.
The $T_d$ and $1T'$ phases have slightly different Poisson ratios for the c-axis response to the strain along a- and b-axes. Since the main difference between these two phases are the stacking of the planes which is controlled by vdW interactions, this small difference in Poisson ratio is also expected.

\subsection{Electron-phonon Coupling in the $T_d$ and $1T'$ phases}

\subsubsection{A single layer MoTe$_2$}

In order to get a better understanding of the origin of superconductivity in MoTe$_2$, we first study an individual MoTe$_2$ layer. We consider both the centrosymmetric $1T'$-phase and the
non-centrosymmetric $T_d$-phase. The single layer of MoTe$_2$ is obtained by deleting the 2nd
layer in the unit cell of $1T'$- and $T_d$-phases and taking the c-axis as $16~\AA$, which gives enough vacuum between the layers. Table~\ref{T2Doptstr} shows the initial 
symmetry of the layer obtained from $T_d$-phase. 
After structural optimizations, we surprisingly discovered that both single layers converge
to the same structure with a symmetry that recovers the inversion symmetry (see Table\ref{T2Doptstr}). We repeat the structural optimization with positive and negative charge doping as well as with different strains in $ab-$plane but the optimization always yield a centrosymmetric
configuration. Hence, we conclude that the origin of the non-centrosymmetric phase of the $T_d$-structure is due to weak inter-layer interactions; Once the inter-planes are separated, there is
no reason for the system to keep the non-centrosymmetric phase and the systems recover inversion
symmetry. This is rather an unexpected and important findings as lack of inversion symmetry is a requirement for the Weyl state and for topological superconductivity.  

\begin{table}
	\caption{\label{T2Doptstr}
Initial atomic positions and symmetry of a single layer MoTe$_2$ obtained from
$1T'-$ and $T_d-$phases, respectively. The final optimized structure
is shown at the bottom. Note that the optimized structure has the inversion
symmetry.  }
	\begin{ruledtabular}
		\begin{tabular}{lcc}
Initial Structure from $T_d$	& P1m1 (\#6): 	& 	 	 \\
a=6.3818 \AA, 	b=3.5042 \AA	&    (x,y,z)      &  (x,-y,z)      \\
			\hline
			Mo1	1a\hfill	 0.5968	& 0	& 0.5004 \\
			Mo2 	1b\hfill	 -0.0403& 0.5	&0.5150\\
			Te1  	 1a\hfill	 0.8536	& 0	&0.6565\\
			Te2  	 1b\hfill	 0.7001	& 0.5	&0.3589\\
			Te3  	 1a\hfill	 0.2053&  0.	&0.4040\\
			Te4  	 1b\hfill	 0.3514&  0.5	&0.6112\\ \hline
 
 Optimized Structure  	        & 	P 1 21/m 1 (\#11):                 		     & \\
a=6.3818 \AA, 		&    (x,y,z)      &  (-x,y+1/2,-z)      \\
 b=3.5042 \AA	        &    (-x,-y,-z)   &  (x,-y+1/2,z)       \\
			\hline
			Mo1	2e\hfill 0.31925	& 0.25	& 0.49401\\
			Te1  	2e\hfill 0.57863	& 0.25	& 0.62955\\
			Te2  	2e\hfill-0.07014	& 0.25	& 0.40736\\
		\end{tabular}
		 
	\end{ruledtabular}
\end{table}

\begin{figure}
\includegraphics[scale=0.6]{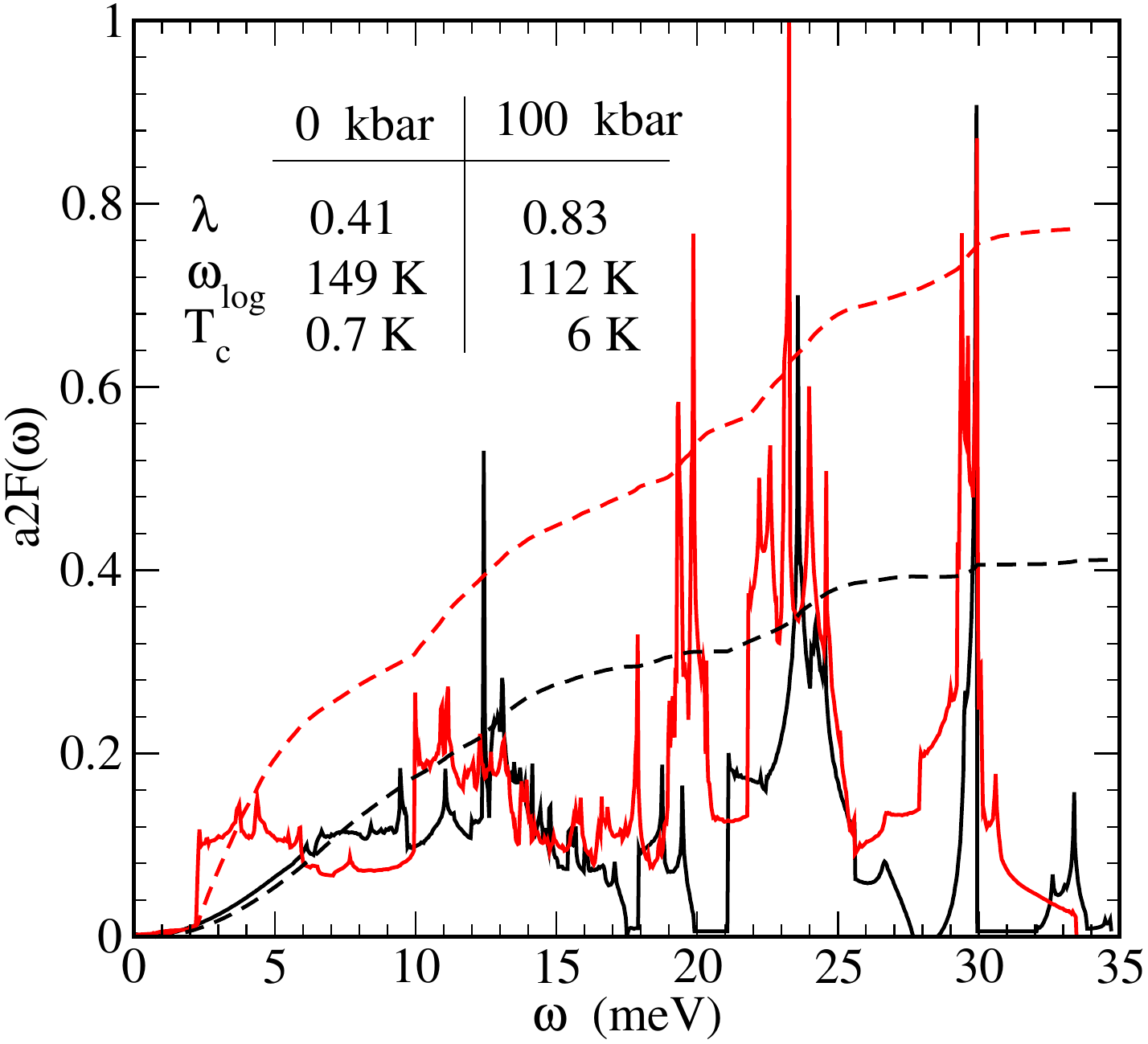}
\caption{\label{T2D_a2F_0kbar_100kbar} Eliashberg function $a^{2}F(\omega)$ and total 
$\lambda$ (dashed line) for a single MoTe$_2$ layer at zero (black) and 100 kbar (red) pressures, respectively. The inset summarizes the total el-ph coupling $\lambda$, logarithmic average phonon frequency and the superconducting temperature T$_c$ (K) for $\mu^{*}=0.1$.  
}
\end{figure}

Next, we study phonons and the electron-phonon (el-ph) coupling in single layer MoTe$_2$ for zero 
and 100 kbar (i.e. 10 GPa) external pressure (within ab-plane). The phonon spectrum and
the electron-phonon couplings are summarized in  
Figure~\ref{T2D_a2F_0kbar_100kbar}. At zero external pressure, we obtained el-ph
coupling $\lambda ~ 0.41$ and $\omega_\mathrm{log}~ 149 K$, which yields a $T_c$ of
0.83 K for $\mu^* = 0.1$. With applied pressure of 100 kbar (i.e. 10 GPa), the
$\lambda$ increases almost twice while $\omega_\mathrm{log}$ reduces to 112 K while
the T$_c$ increases to 6 K, which is similar to what we observe experimentally. 
As we shall discuss later, we do not see this trend for the case of bulk MoTe$_2$ 	
in either  of $1T'$ and $T_d$ phases.  From the projected Eliashberg function $a2F$,
we determined that both $Mo$ and $Te$ contribution  equally to $\lambda$. Similarly,
phonon projections along $a-$, $b-$, and $c-axis$ gives similar contributions to 
$\lambda$, indicating that electron-phonon coupling in MoTe$_2$ 	is rather isotropic,
getting contribution from all phonons based on $Mo$ and $Te$ phonons along all three
directions in space. This is a very different situation than in MgB$_2$ where the main
contribution to el-ph coupling comes from limited in-plane B-based phonons. In MoTe$_2$
layer, all phonons at all energies contribute equally to the el-ph coupling.

\begin{figure}
\includegraphics[scale=0.4]{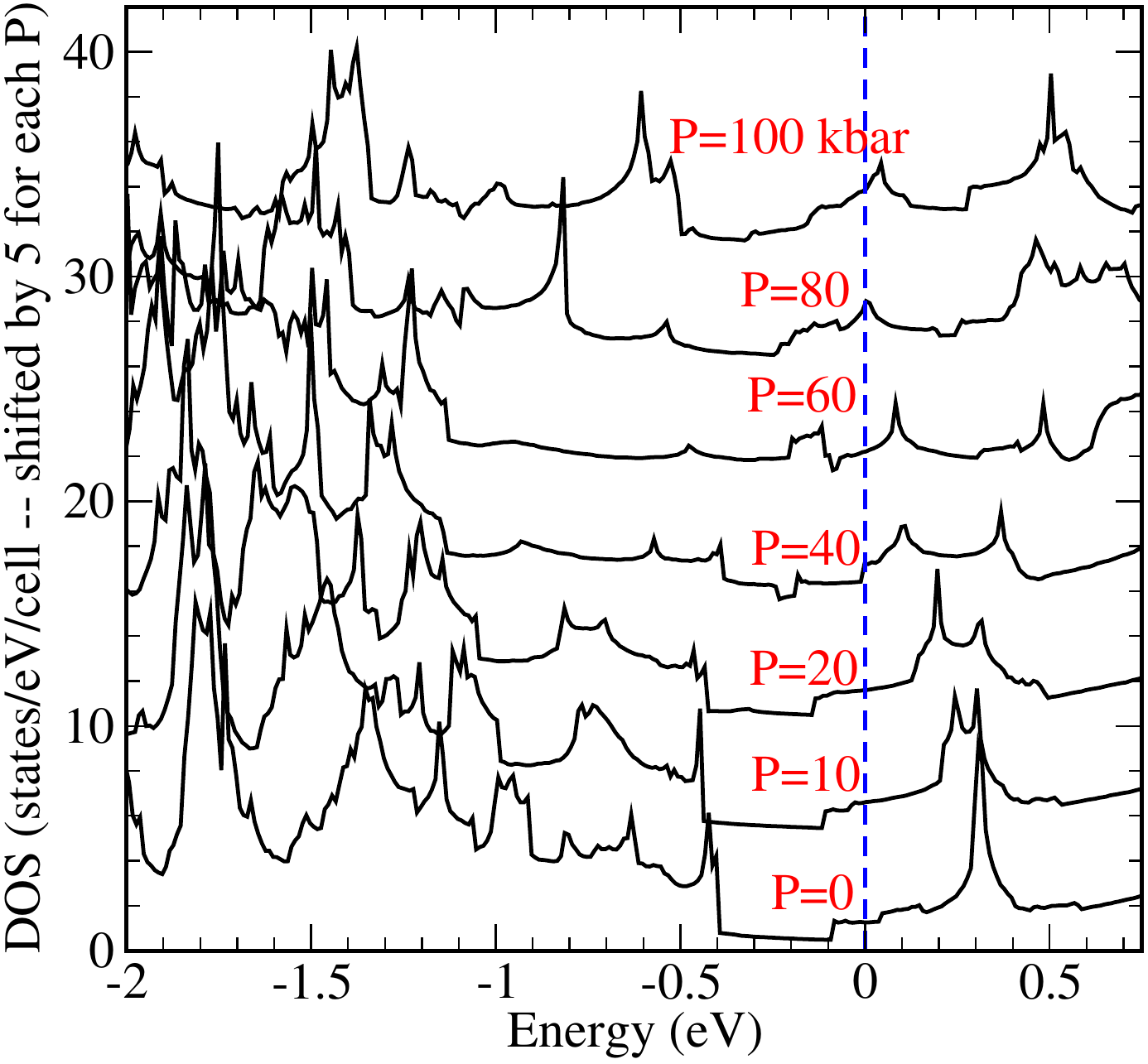}
\caption{\label{T2D_DOS} The electronic density of states (DOS) with
applied pressure for a single layer MoTe$_2$. 
}
\end{figure}

\begin{figure}
\includegraphics[scale=0.4]{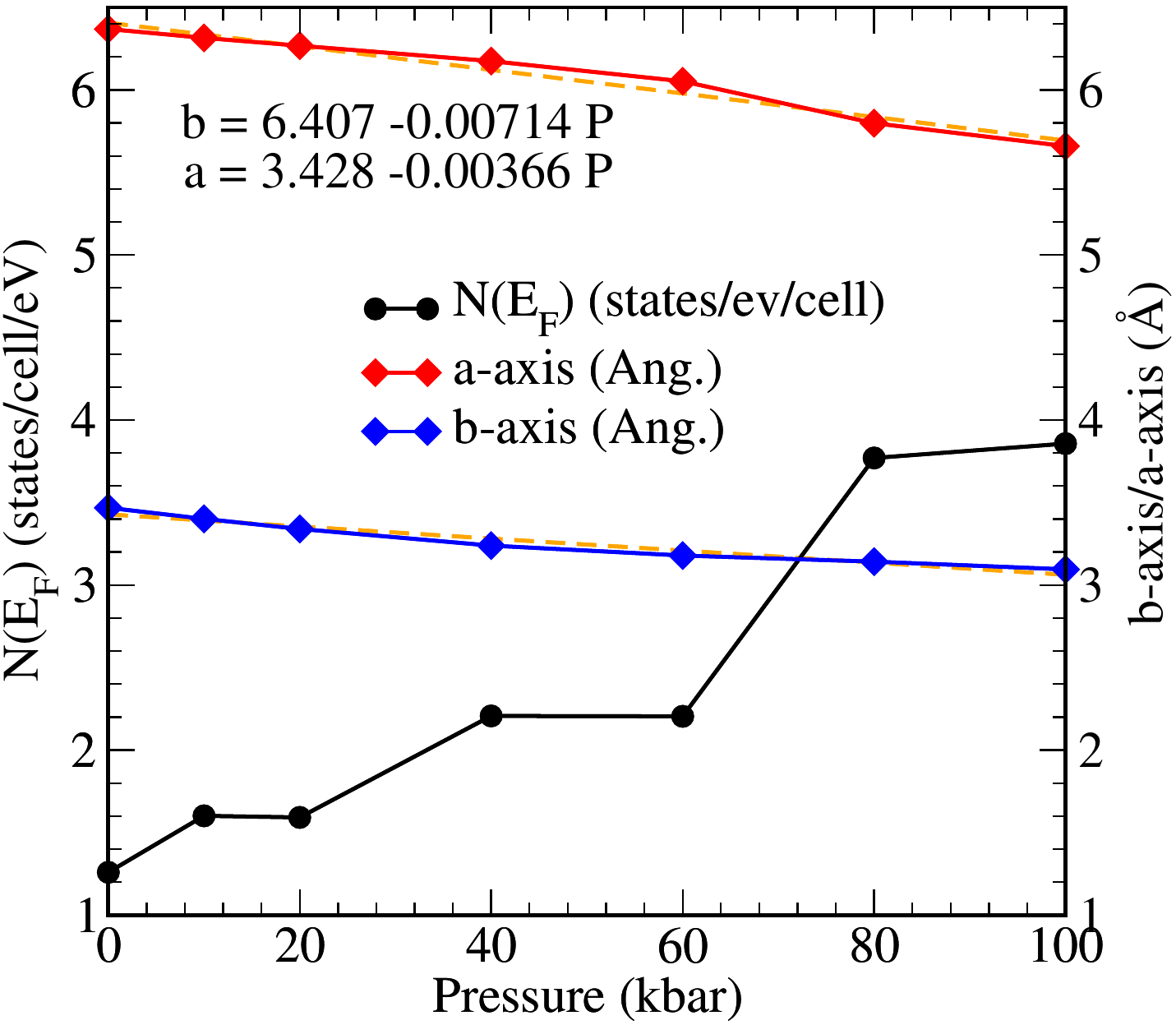}
\caption{\label{T2D_NEF} The lattice parameters and the 
electronic density of states at the Fermi level with applied
in-plane pressure P for single layer MoTe$_2$.    
}
\end{figure}

In order to get better insight to the sharp increase in $T_c$ with applied
pressure, we look at the electronic density of states with pressure which is
shown in Figure \ref{T2D_DOS}.  Some of the states above Fermi level decreases
in energy with applied pressure and eventually intersect the Fermi level near
$P=80$ kbar, causing the increase in the density of states near Fermi level.
This is clearly shown in Figure~\ref{T2D_NEF}. Since the logarithmic phonon
energy is actually decrease with the increasing pressure, the increase in N(E$_F$)
with pressure seems to be the only mechanism to explain the observed increase in
$\lambda$ with applied pressure.

Figure~\ref{T2D_NEF} also shows the lattice parameters $a-$ and $b-$axis with
in-plane applied pressure. Due to single layer nature of MoTe$_2$, the buckling of
the atoms out  of the plane is easier in the case of single layer and therefore
the compressibilty of a- and b-axes are larger than the case of bulk. However
as we shall discuss below, if we take the lattice constants of the 2D layer
at 100 kbar and repeat the el-ph coupling calculations for the case of bulk,
we do not see increase in $\lambda$, which raises quite interesting questions about
the origin of T$_c$ enhancement with pressure observed experimentally.

\subsubsection{Bulk MoTe$_2$: $1T'$ and $T_d$ phases}

\begin{figure}
\includegraphics[scale=0.5]{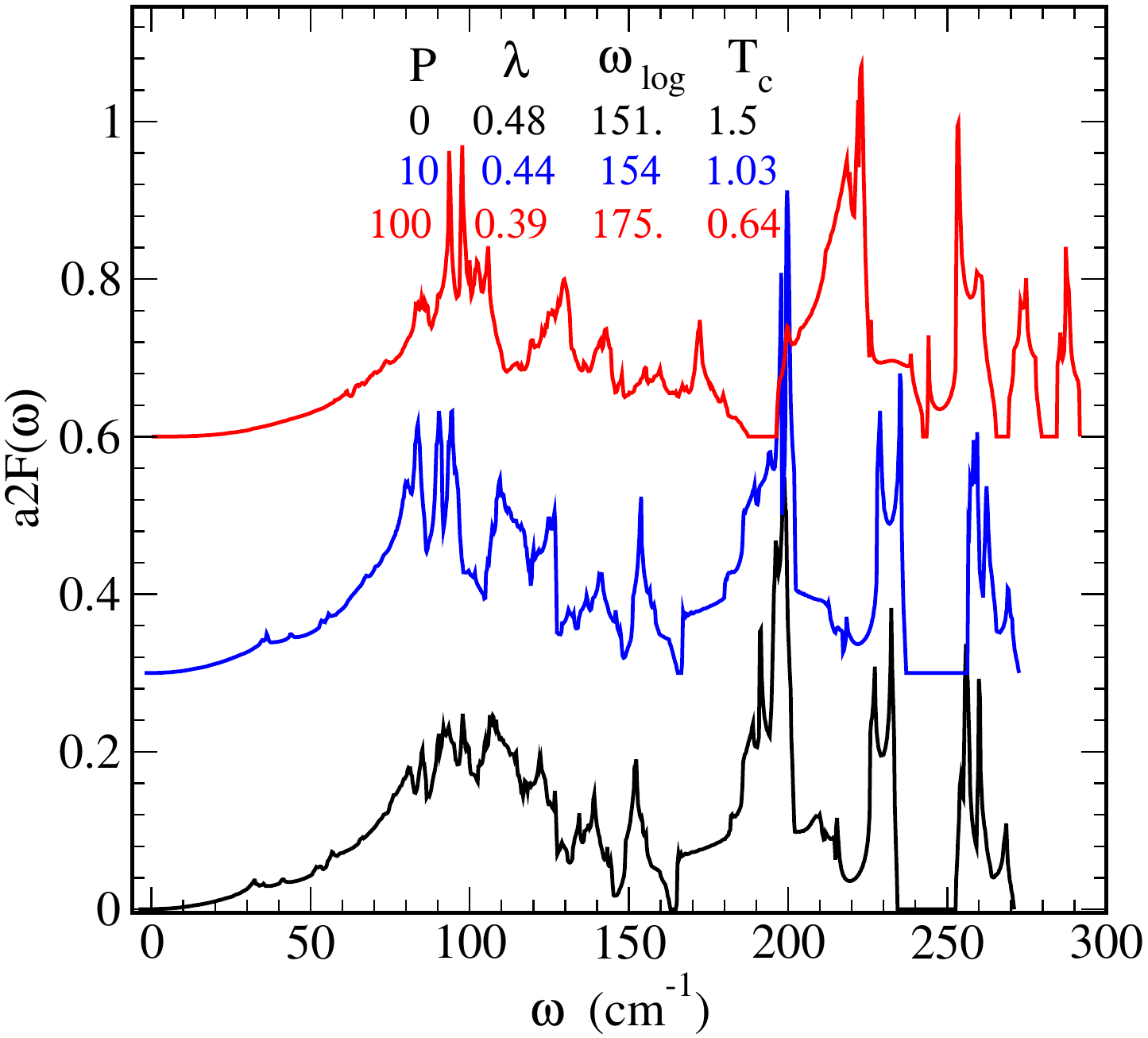}
\includegraphics[scale=0.5]{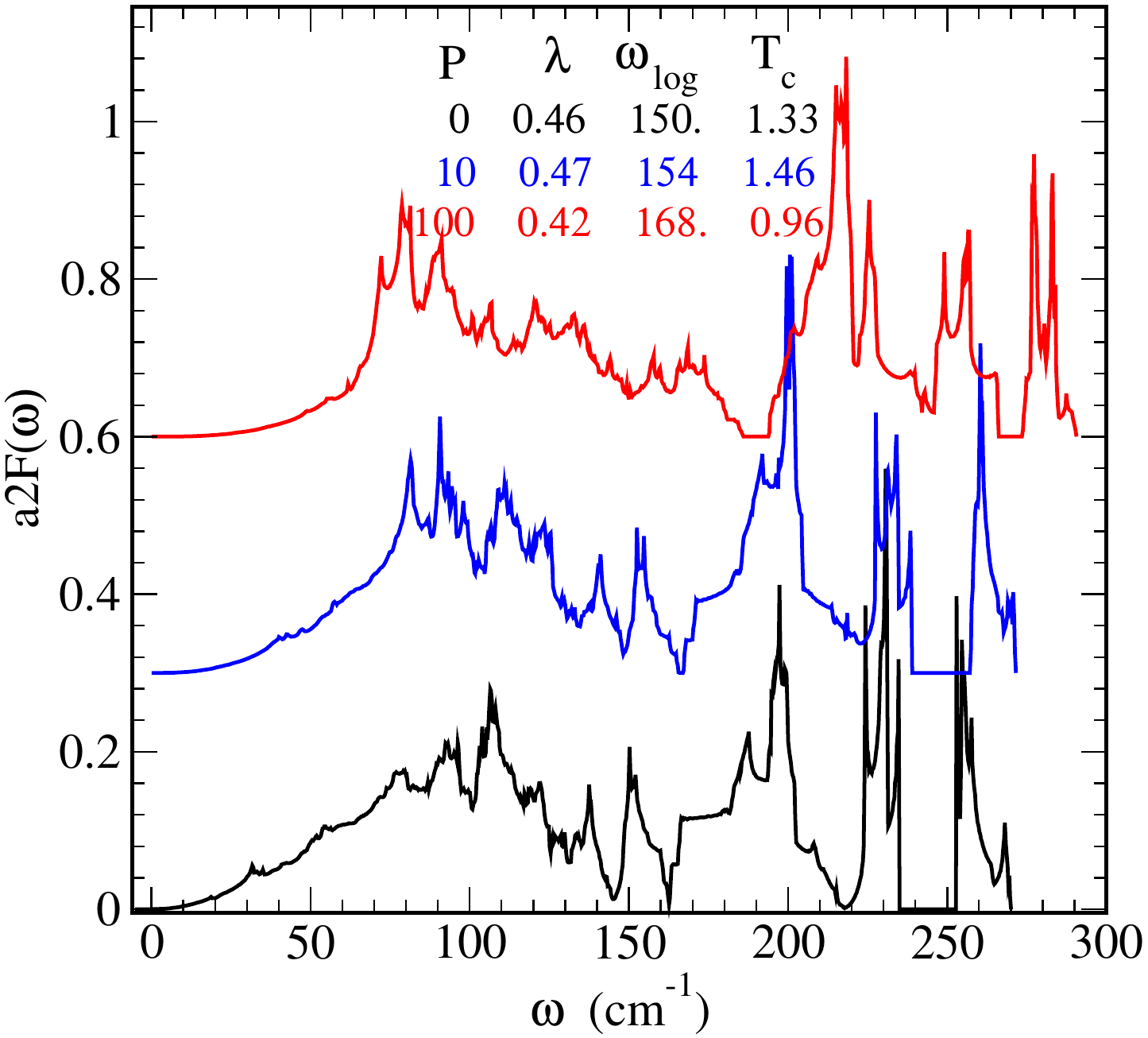}
 
\caption{\label{a2F} Left: Eliasherg function as a function of 
pressure for 
for $T_d$ phase (top) and $1T'$ phase (bottom).
}
\end{figure}

We now discuss the electron phonon coupling in bulk MoTe$_2$ with both phases
at zero and applied pressure. Our results are summarized in Figure~\ref{a2F}. Both phases
show a very similar Eliashberg function at a given pressure with similar 
$\lambda$ and logarithmic average phonon energies. As in the case of single
layer MoTe$_2$, projected $a2F$ shows that the contribution to the total 
$\lambda$ comes from all phonons at all energies and polarizations. The
zero pressure $a2F$ for the bulk phase is quite similar to the isolated
single layer MoTe$_2$, indicating that the main mechanism of superconductivity is
involved within the single MoTe$_2$ plane and the interlayer coupling has no
apparent effect. The main difference between the bulk and the single layer MoTe$_2$
is the pressure dependence of the electron phonon coupling. For the case of
single layer, we found almost an order of magnitude increase in T$_c$ from 
0 kbar to 100 kbar pressure range. However that is not the case for the bulk
MoTe$_2$ in either phases. We did not see significant changes in either
$\lambda$ or $\omega_\mathrm{log}$. Hence based on our calculations,
a conventional phonon-based electron phonon coupling can not explain the
observed enhancement of T$_c$ with pressure. One potential explanation could be
that the applied pressure may decouple the MoTe$_2$ layers and then the decoupled
layers are responsible for the enhanced T$_c$ with pressure, observed experimentally.  Proximity effects may also play a strong role.

\begin{figure}
\includegraphics[scale=0.4]{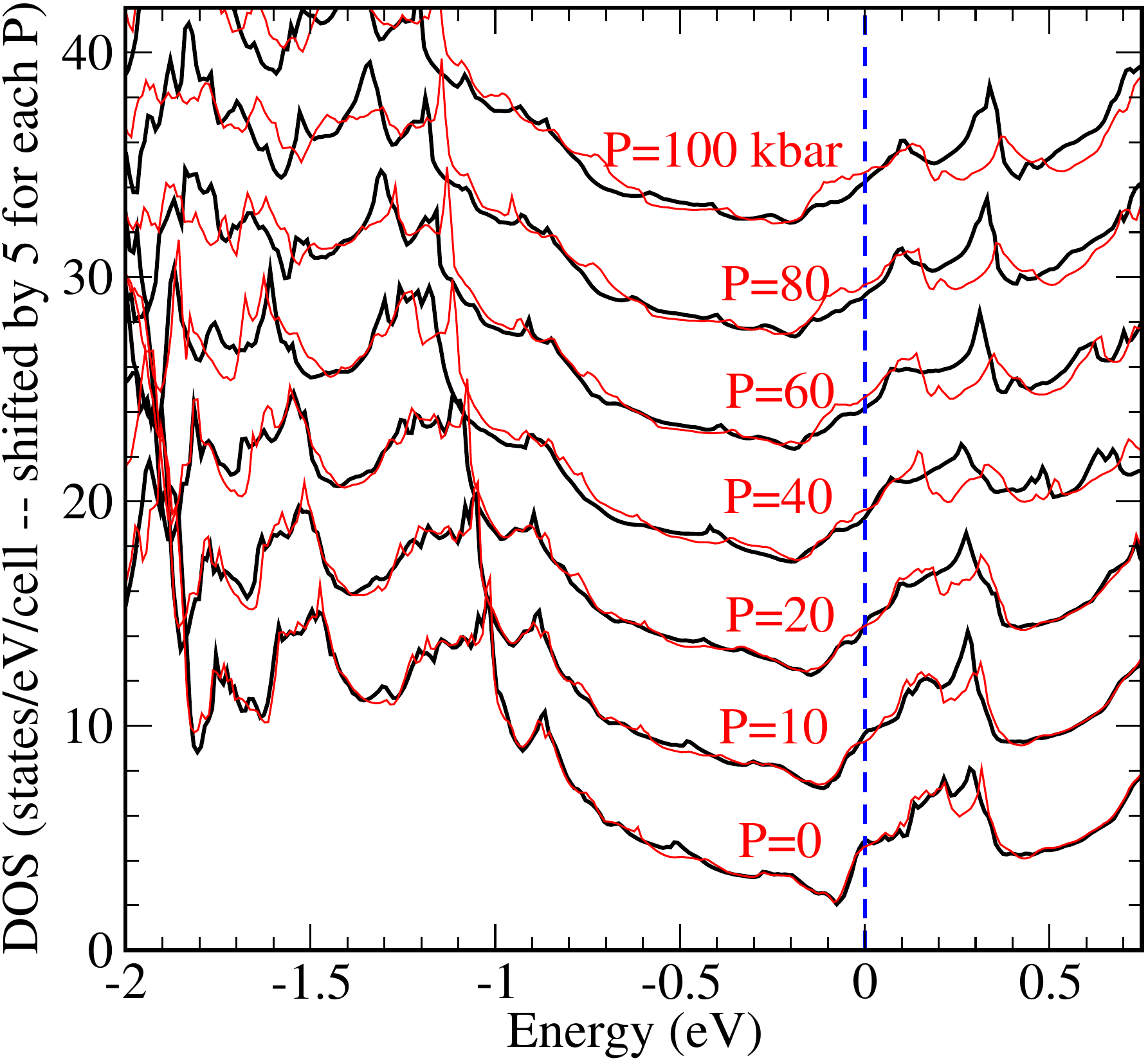}
\includegraphics[scale=0.45]{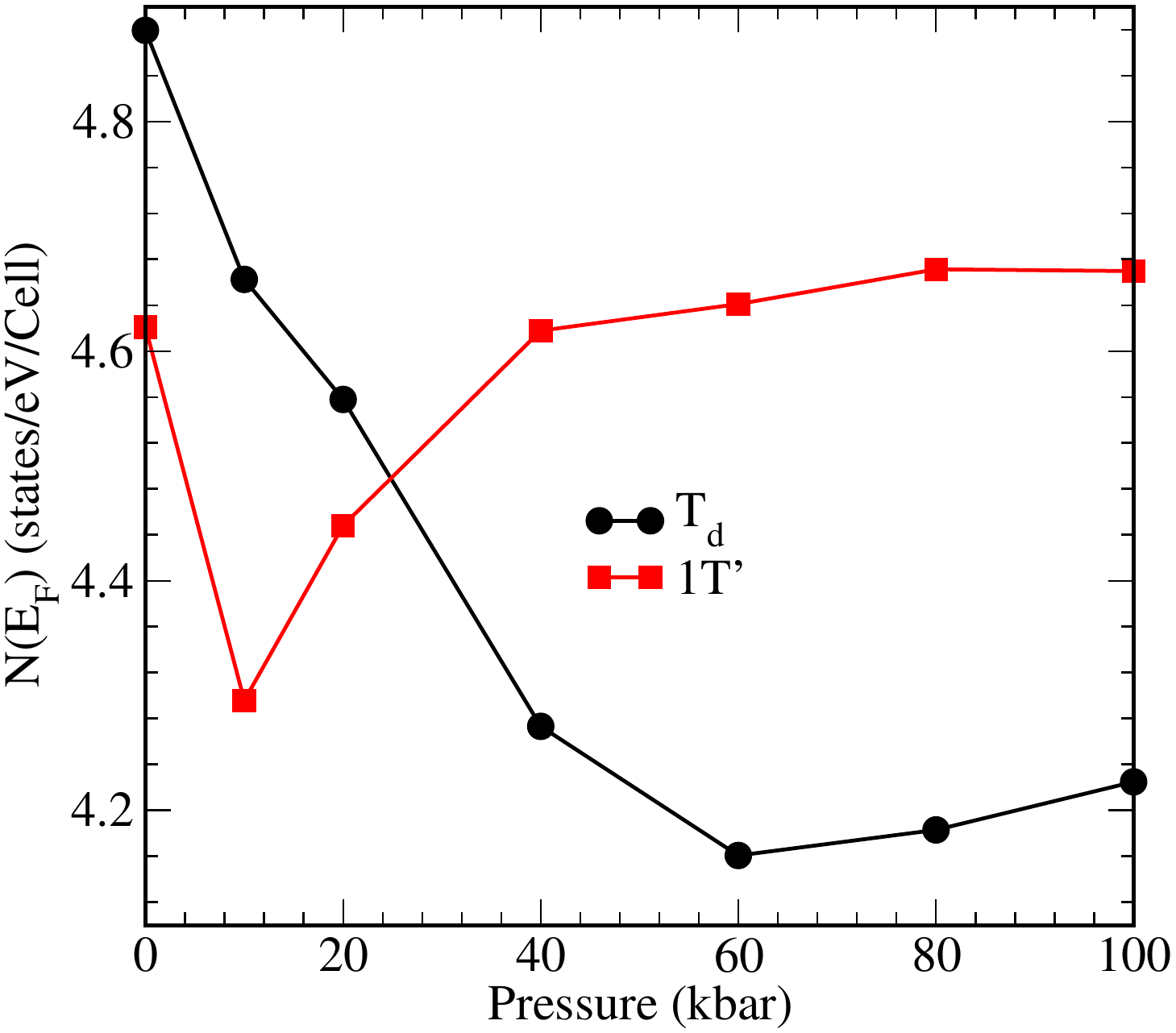}
 
\caption{\label{DOS_P} Top: The total electronic density of
states (DOS) for $T_d$ (black) and $1Tp$ phases (red). 
Bottom: The density of states at the Fermi level, $N(E_F)$,
as a function of pressure for both phases.
}
\end{figure}

In order to have a better understanding of the effect of pressure on the
superconductivity, we study the density of states as a function of pressure
for both phases. In the case of single layer, we observed that the density
of states at the Fermi level, $N(E_F)$ is the main factor that increases with
pressure and in turn, yields higher el-ph coupling constants at high pressures.
As shown in Figure~\ref{DOS_P}, we do not see a similar pressure dependence
in the bulk case. The $N(E_F)$ actually decreases slightly with applied pressure
for the case of $T_d$ phase, which makes sense as the lattice parameters
decrease, the orbital overlap increases, giving rise to wider band dispersion
lowering $N(E_F)$. The behavior is slightly different for the $1T'$ phase
where the density of states first decreases and then increases and stay constant
with pressure. We attribute this behavior to an increase of the cell angle $\beta$
which increases by $3^o$ with applied pressure. In any case, we do not see
any significant increase in the density of states with pressure as we found
in the case of single layer MoTe$_2$. The main effect of pressure 
is a slight increase of the logarithmic
average phonon energies but the increase is not large enough to explain
the large increase in T$_c$ with only 10 GPa pressure as experimentally
observed.

\begin{figure}
\includegraphics[scale=0.5]{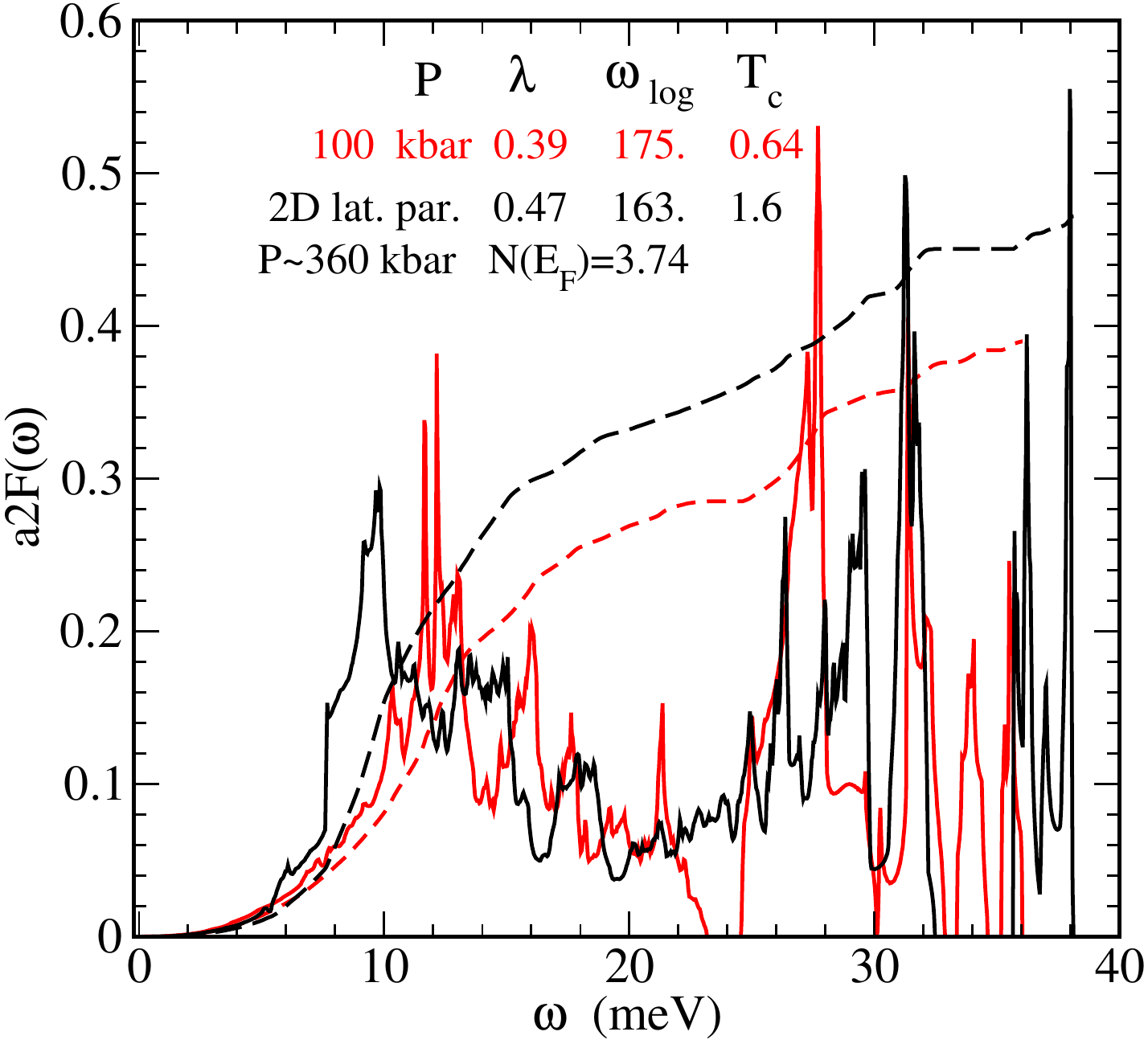}
\caption{\label{3Dat2Dlatpar} Comparison of Eliashberg function
$a2F$ for the $T_d$ phase at P=100 kbar and at the 
lattice parameters of single layer MoTe$_2$ at 100 kbar
(which roughly corresponds to 360 kbar). 
}
\end{figure}
One may wonder if we use the lattice parameters of the 2D MoTe$_2$ at 100 bar
for the case of bulk MoTe$_2$, can we get enhanced $\lambda$ and therefore 
a higher T$_c$ with applied pressure. For the case of single layer MoTe$_2$, 
the buckling of the atoms is easier and therefore a- and b-axes change
almost twice as fast as in the bulk case. Indeed if we use the lattice
parameters of the 2D structure at 100 kbar, we obtained pressures about
360 kbar for the bulk case. Figure~\ref{3Dat2Dlatpar} summarizes the
results for the Eliashberg function and other parameters. Interestingly
despite using the same lattice constants as single layer MoTe$_2$, we do 
not get either higher $N(E_F)$ or larger $\lambda$. The results are
rather similar to the case of 100 kbar, which was also shown in the
same figure for comparison. Hence, we can not explain the observed high
T$_c$ with pressure by taking compressed lattice parameters, even that of
~360 kbar.  We clearly need more study to fully understand the origin of
the pressure dependence of the superconductivity observed in MoTe$_2$ but
based on our results it seems that maybe pressure acts to decouple layers of MoTe$_2$ which is
responsible for the observed behavior. More study, in particular, more experiments
to test these ideas would be quite interesting.

\begin{figure}
\includegraphics[scale=0.5]{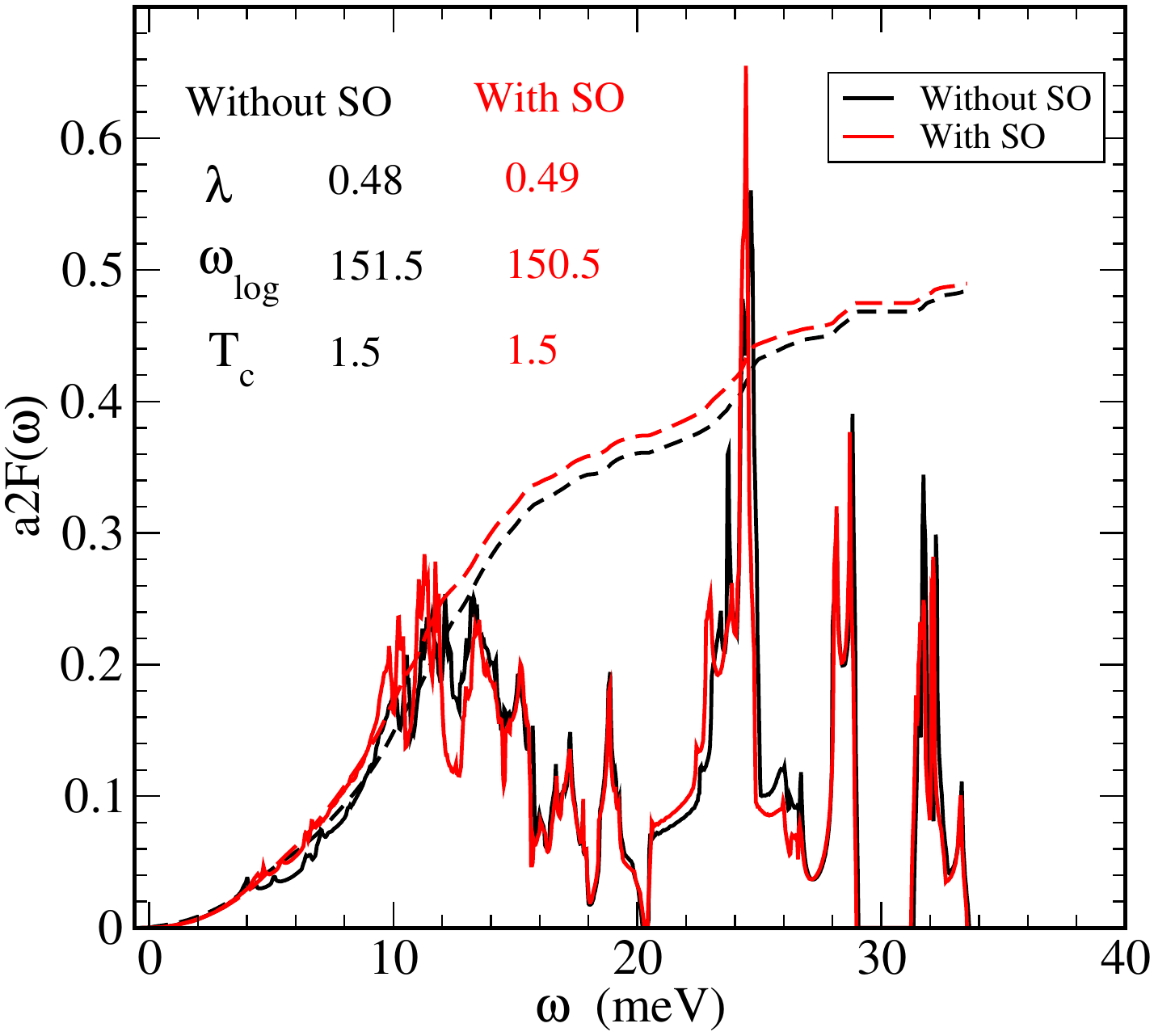}
\caption{\label{SO} Comparison of Eliashberg function
$a2F$ for the $T_d$ phase with and without spin-orbit interaction. 
}
\end{figure}

Finally, we study the effect of spin-orbit interactions on the phonon spectrum and
the electron-phonon coupling for the $T_d$-phase. Due to lack of inversion in the 
$T_d$-phase, we thought that SO interaction may be important but as shown in Figure~\ref{SO}, the
effect of SO both on the phonon spectrum and the el-ph coupling is rather small.
We also checked the effect of SO-interaction in the $1T'$-phase at particular q-points
and similar to $T_d$-phase, we did not see significant change with SO interaction turned on.
Hence, it seems to be safe to ignore the SO-interactions in structure and lattice dynamics of
MoTe$_2$ in either phases.

\section{Structure Determination}
\subsection{Structure Notation}
The standard (hkl) notation used to describe reciprocal lattice planes in space group 11 (1T' phase) and 31 (T$_d$ phase) uses different conventions for the labeling of the crystallographic a,b, and c unit cell directions. Here we will use the monoclinic convention where c is the long axis along the van de Waals bonding direction with a lattice constant of order 13.8 $\AA$, a is the axis opposite the monoclinic distortion with lattice constant of order 6.3 $\AA$, and b is the remaining axis with lattice constant of order 3.5 $\AA$. The orthorhombic phase is effectively the same as the monoclinic angle but with a $\beta$ angle of 90$^{0}$, which effectively transposes the h and k labels from the standard orthorhombic notation in space group Pmn2$_1$ (this can be alternatively described as using the non-standard Pnm2$_1$ space group in Hermann-Mauguin notation). \par

As our metric for structure we focused on the reflections from the (201) like lattice planes. In the monoclinic phase the (201) and the (20$\bar{1}$) reflections split in reciprocal space, while there is just once centrally located (201) reflection in the orthorhombic phase. 

We mounted single crystals from the same synthesis batch as our transport measurements in the (h0L)$_M$ zone and monitored the (00L)$_M$ reflections, the (201)$_M$ reflections, and the (20$\bar{1}$)$_M$ reflections. In this way, we could monitor changes in the c-axis as well as the merging of the (201)$_M$ and (20$\bar{1}$)$_M$ reflections into a single (201)$_M$ reflection as the crystal transitioned from the 1T' to T$_d$ phase. Due to monoclinic twinning, we can best monitor the phase fraction of the two phases by tracking the integrated intensity of the (201)$_M$ reflection at the reciprocal space position for the orthorhombic phase. Details of the neutron scattering measurements can be found in the methods section. \par

\subsection{Structural refinements}
Powder neutron measurements were performed on the BT-1 diffractometers at NIST using the Cu(311) monochromator option at 60' collimation and 1.540 $\AA$. Powder patterns were fit using the Reitveld method in GSAS-II to obtain lattice constants, atomic positions, and thermal parameters. The crystalographic parameters for powder data taken at 300 K and 3 K are shown in table  S2 while atomic positions for the two refined structures are shown in tables S3 and S4. We see a small fraction of metallic Te as an impurity phase, which is included in the refinement as a secondary phase. Due to the anisotropic cleaving of MoTe$_2$, we see expected prefered orientation in the powder which is treated as a 4th order spherical harmonic. \par

 Single crystal data was also taken with the HB-3A 4-circle single crystal diffractometer at ORNL using  using a 34.22 meV neutron beam. Lattice parameters at a range of pressures and temperatures were calculated using the 4-circle data reduction package within MANTID from the ORNL data. These parameters are shown in table S5. Strain broadening in the high pressure state and a large sample environment background from our pressure cell prevented us from refining atomic positions beyond our powder refinements, though lattice parameters are consistent with an altered a/b ratio for the 1.5 GPa state. \par

\par

\begin{table}
\centering
\begin{tabular}{ |P{3cm}||P{3cm}|P{3cm}|}
 \hline
 \multicolumn{3}{|c|}{Neutron Powder Diffraction Extracted Crystallographic} \\  \multicolumn{3}{|c|}{Parameters and Refinement Statistics}\\
 \hline
 Temperature & 300 K & 3 K\\
 Space Group & \textit{P2$_{1}$/m}, (No. 11) & \textit{Pmn2$_{1}$}, (No. 31) \\
 a ($\AA$)   & 6.3281(3)           & 3.46464(13)  \\
 b ($\AA$)   & 3.47703(17)         & 6.30716(23)  \\
 c ($\AA$)   & 13.8182(11)         & 13.8431(6)   \\
 $\alpha$    & 90$^{\circ}$        & 90$^{\circ}$ \\
 $\beta$     & 93.882(5)$^{\circ}$ & 90$^{\circ}$ \\
 $\gamma$    & 90$^{\circ}$        & 90$^{\circ}$ \\
 Cell Volume & 303.346(20)         & 302.500(13)  \\
 $\chi^2$    & 1.01                & 1.25 \\
 wR  (\%)    & 7.00                & 7.68 \\

 \hline

\end{tabular}
\caption{Powder structure determination results and refinement statistics at 300 K and 3 K for quenched MoTe$_2$ }
\end {table}

\begin{table}
\centering
\begin{tabular}{ |P{1cm}|P{1.8cm}|P{1.8cm}|P{1.8cm}|p{1.8cm}|P{1.4cm}|  }
\hline
 \multicolumn{6}{|c|}{300 K \textit{P2$_1$/m} Phase Atomic positions} \\
\hline
Atom &   X &   Y &    Z &     Uiso &    Wyckoff Position \\
Mo$_1$ & 0.1828(7)  &   0.25 &   0.0083(4) &    0.0052(10) & 2e  \\ 
Mo$_2$ & 0.3194(8)  &  -0.25 &   0.5062(4) &    0.0087(10) & 2e  \\
Te$_1$ & 0.5880(10) &   0.25 &   0.1064(5) &    0.0097(13) & 2e  \\ 
Te$_2$ & 0.0966(9)  &  -0.25 &   0.1493(5) &    0.0102(14) & 2e  \\ 
Te$_3$ & 0.5571(10) &  -0.25 &   0.3513(5) &    0.0108(14) & 2e  \\ 
Te$_4$ & 0.0563(10) &   0.25 &   0.3953(4) &    0.0110(13) & 2e  \\ 

\hline
\end{tabular}
\caption{Refined structural parameters for 1T' MoTe$_2$ at 300 K  }
\end {table}

\begin{table}
\centering
\begin{tabular}{ |P{1cm}|P{1.8cm}|P{1.8cm}|P{1.8cm}|P{1.4cm}|  }
\hline
 \multicolumn{5}{|c|}{3 K \textit{Pmn2$_1$} Phase Atomic positions} \\
\hline
Atom &   X &   Y &    Z &           Wyckoff Position \\
Mo$_1$ &  0.0 &   0.6061(7)  &   0.497244 &        2a \\ 
Mo$_2$ &  0.0 &   0.0293(8)  &   0.014242 &        2a \\
Te$_1$ &  0.0 &   0.8659(10) &   0.653546 &        2a \\  
Te$_2$ &  0.0 &   0.6411(9)  &   0.112020 &        2a \\ 
Te$_3$ &  0.0 &   0.2877(11) &   0.857259 &        2a \\  
Te$_4$ &  0.0 &   0.2147(10) &   0.401510 &        2a \\ 

\hline
\end{tabular}
\caption{Refined structural parameters for T$_d$ MoTe$_2$ at 3 K. Unsurprisingly, the additional refinement of thermal parameters at 3 K did not improve the overall refinement and are thus not included here. }
\end {table}
\begin{table}
\begin{tabular}{ |p{1cm}||p{2cm}|p{2cm}|p{2cm}|  }
 \hline
 \multicolumn{4}{|c|}{Single Crystal Lattice Parameters} \\
 \hline
 & 5 K \newline Ambient Pressure & 5 K \newline 1.5 GPa&240 K \newline 1.5 GPa\\
 \hline
 a ($\AA$) & 6.31$\pm$ 0.027 &6.27$\pm$ 0.013&6.30$\pm$ 0.027\\
 b ($\AA$)&3.47$\pm$ 0.023  &3.44$\pm$ 0.013   &3.476$\pm$ 0.025 \\
 c ($\AA$)& 13.93$\pm$ 0.044 & 13.71$\pm$ 0.019&13.63$\pm$ 0.034\\
 $\alpha$ & 90$^{\circ}$ & 90$^{\circ}$ & 90$^{\circ}$\\
 $\beta$ &90$^{\circ}$ & 90$^{\circ}$ & 94.76$^{\circ}\pm$ 0.29$^{\circ}$\\
 $\gamma$ & 90$^{\circ}$ & 90$^{\circ}$ & 90$^{\circ}$\\
 
 \hline
 
\end{tabular}
\caption{Unit cells parameters fit from 4-circle measurements on HB-3A at ORNL. Lattice parameters are calculated using MANTID's 4-circle data reduction package and UB-matrix calculator. }
\end {table}

\bibliography{MoTe2,sup}

\end{document}